\documentclass[11pt,a4paper]{article}
\pdfoutput=1
\usepackage{jcappub}

\usepackage{xcolor}

\usepackage{bm}
\usepackage{subfigure}
\usepackage{feynmp}
\usepackage{extarrows}
\usepackage{slashed}
\usepackage{dcolumn}
\usepackage{verbatim}
\usepackage{here}
\usepackage{multirow} 
\allowdisplaybreaks[4]

\numberwithin{equation}{section}

\newcommand{\FR}[2]{\displaystyle\frac{\,{#1}\,}{#2}}
\newcommand{\fr}[2]{\mbox{$\frac{\,{#1}\,}{#2}$}}
\newcommand{\n}{\nonumber}

\def\bge{\begin{equation}}
\def\ede{\end{equation}}
\def\bga{\begin{aligned}}
\def\eda{\end{aligned}}
\def\bgp{\begin{pmatrix}}
\def\edp{\end{pmatrix}}
\def\bgs{\begin{subequations}}
\def\eds{\end{subequations}}
\newcommand{\beq}{\begin{equation}}
\newcommand{\eeq}{\end{equation}}
\newcommand{\bq}{\begin{equation}}
\newcommand{\eq}{\end{equation}}
\newcommand{\ba}{\begin{array}}
\newcommand{\ea}{\end{array}}
\newcommand{\beqa}{\begin{eqnarray}}
\newcommand{\eeqa}{\end{eqnarray}}
\newcommand{\beqs}{\begin{subequations}}
\newcommand{\eeqs}{\end{subequations}}

\newcommand{\order}[1]{\mathcal{O}({#1})}
\def\di{{\mathrm{d}}}
\def\D{{\mathrm{D}}}

\def\[{\left[}
\def\]{\right]}
\def\({\left(}
\def\){\right)}

\def\ra{\rightarrow}

\def\T{\mathcal{T}}

\def\pd{\partial}
\def\ld{{\mathcal{L}}}

\def\la{\langle}\def\ra{\rangle}
\def\sla{\slashed}

\def\to{\rightarrow}

\def\ii{\mathrm{i}}

\def\al{\alpha}
\def\be{\beta}
\def\ga{\gamma}
\def\de{\delta}
\def\ep{\epsilon}
\def\ka{\kappa}
\def\lam{\lambda}
\def\rh{\rho}
\def\si{\sigma}

\def\phib{\bar{\phi}}

\def\Mp{M_{\mathrm{Pl}}}

\def\cut{\Lambda_{\text{UV}}}

\def\End{\end{document}}

\newcommand{\overbar}[1]{\mkern 1.5mu\overline{\mkern-1.5mu#1\mkern-1.5mu}\mkern 1.5mu}

\title{\huge Higgs Gravitational Interaction, Weak Boson Scattering, and Higgs Inflation in Jordan and Einstein Frames}

\author[a]{\large Jing Ren,}
\author[a,b]{\large~Zhong-Zhi Xianyu,}
\author[a,c,d]{\large~Hong-Jian He\,}

\affiliation[a]{Institute of Modern Physics and Center for High Energy Physics,\\
                Tsinghua University, Beijing 100084, China}
\affiliation[b]{Theoretical Particle Physics and Cosmology Group, Department of Physics,\\
                King's College London, London WC2R 2LS, UK}
\affiliation[c]{Center for High Energy Physics, Peking University, Beijing 100871, China}
\affiliation[d]{Kavli Institute for Theoretical Physics China, CAS, Beijing 100190, China}

\emailAdd{jingren2004@gmail.com, xianyuzhongzhi@gmail.com, hjhe@tsinghua.edu.cn}

\abstract{
\\
 We study gravitational interaction of Higgs boson through the unique dimension-4
 operator $\,\xi H^\dag H\mathcal{R}$\,,\, with $\,H\,$ the Higgs doublet and
 $\,\mathcal{R}\,$ the Ricci scalar curvature.  We analyze the effect of this
 dimensionless nonminimal coupling $\,\xi\,$ on weak gauge boson scattering in both
 Jordan and Einstein frames. We explicitly establish the longitudinal-Goldstone boson
 equivalence theorem with nonzero $\xi$ coupling in both frames, and analyze the
 unitarity constraints.
 We study the $\xi$-induced weak boson scattering cross sections at ${\cal O}(1-30)$TeV scales,
 and propose to probe the Higgs-gravity coupling via weak boson scattering experiments
 at the LHC\,(14\,TeV) and the next generation $pp$ colliders ($50-100$\,TeV).
 We further extend our study to Higgs inflation, and quantitatively derive
 the perturbative unitarity bounds via coupled channel analysis, under
 large field background at the inflation scale. We analyze the unitarity constraints
 on the parameter space in both the conventional Higgs inflation and the improved models
 in light of the recent BICEP2 data.
}

\keywords{\\
Particle physics\,-\,Cosmology connection, Inflation, Quantum gravity phenomenology,  \\
Cosmology of theories beyond the SM
\\[5mm] 
JCAP (2014), in Press [\,arXiv:1404.4627 [gr-qc]\,]. \hfill {KCL-PH-TH/2014-25}
}

\begin{document}

\maketitle

\setlength{\baselineskip}{18pt}

\setcounter{page}{2}
\vspace*{10mm}
\section{Introduction}
\label{intro}
\vspace*{2mm}

The LHC discovery of a Higgs boson ($\sim\! 125$\,GeV) \cite{LHC2012,LHC2013} has begun a new era
for particle physics and its interface with cosmology and gravitation research.
The Higgs boson plays a distinctive role in the standard model (SM) of particle physics.
It is the only fundamental scalar particle in the SM.
It spontaneously breaks the electroweak gauge symmetry and provides
the origin of inertial masses for all massive particles:
weak gauge bosons, quarks, leptons and neutrinos.
As measured from the current LHC experiments \cite{LHC2013}, the properties of this 125\,GeV
new particle (including its signal strengths in several decay channels and its spin/parity)
are still compatible with the SM predictions. However, most of its couplings
(especially its Yukawa couplings and self-couplings) have not yet been tested,
and possible new physics can cause deviations of these couplings from those of
the SM Higgs boson, which will be probed at the upcoming LHC runs and the future
$e^+e^-$ and $pp$ colliders \cite{peskin}.

Furthermore, the SM is apparently incomplete for not containing the gravitational force,
despite all SM particles join in gravitation. Even though the Einstein general relativity (GR)
still gives the best description of gravitation, it is a notoriously nonrenormalizable
field theory \cite{HV}. Hence, it is compelling to incorporate the SM and GR together as
a joint low energy effective theory below the Planck scale, and explore the testable
effects from this unavoidable interface.
Especially, with the LHC Higgs discovery \cite{LHC2012,LHC2013}, we are strongly motivated to
study gravitational interactions of the Higgs boson, because the Higgs boson generates
inertial masses for all SM particles while the gravity force arises from their
gravitational masses.

Such an effective theory always has an ultraviolet (UV) cutoff, at or below the
Planck mass. In this formulation, one can write down the most general action under
all known symmetries, as a series of effective operators with increasing mass-dimensions
and with proper suppressions by the UV cutoff \cite{EFT}.
Thus, for the experimentally accessible energy ranges being well below the UV cutoff,
the leading terms of this effective action can provide
a fairly good approximation of the full theory.

In fact, under the $\,U(1)_Y^{}\otimes SU(2)_L^{}\otimes SU(3)_c^{}\,$ gauge symmetries
and given the three families of leptons and quarks,
the SM Lagrangian is already the most general effective theory up to dimension-4 operators.
On the other hand, the Einstein-Hilbert action of general relativity represents the leading
terms of mass-dimension zero and two, under the generally covariant expansion,
\bge
  S_{\text{EH}}^{} ~=~
  \Mp^2\int\!\!\di^4x\,\sqrt{-g}\,\Big(\!-\Lambda+\FR{1}{2}\mathcal{R}\Big),
\ede
where  $\,\Mp=(8\pi G)^{-1/2}\simeq 2.44\times 10^{18}$\,GeV gives the reduced Planck mass,
$\,\Lambda$\, denotes the cosmological constant, and $\,\mathcal{R}$\, is the Ricci scalar.
One can continue to write down more operators with higher mass-dimensions in this series.
Up to dimension-4, we have
\bge
S^{}_{\text{G}4} ~=\,
\int\!\!\di^4x\,\sqrt{-g}\,
\(\,c_1\mathcal{R}^2+c_2\mathcal{R}_{\mu\nu}\mathcal{R}^{\mu\nu}\,\)\,.
\ede
There is another possible term $\,\mathcal{R}_{\mu\nu\rh\si}\mathcal{R}^{\mu\nu\rh\si}$\,
allowed by the symmetry, but is not independent up to integration by parts.

In the GR, the gravitational interactions of matter fields (including all the SM particles)
are introduced in the manner of minimal coupling.
In this way, the fluctuation of spacetime metric
couples to matter fields through the energy-momentum tensor,
\beqa
\label{eq:MC}
S_{\text{MC}} \,=\,
-\int\!\!\di^4x\,\sqrt{-g}\,\de g_{\mu\nu} T^{\mu\nu}_{\text{SM}} \,,
\eeqa
where the energy-momentum tensor $\,T_{\text{SM}}^{\mu\nu}$\, contains the SM fields.
For bosonic fields, the rule of minimal coupling is practically equivalent to
making two replacements in the SM Lagrangians.
One is to replace Minkowski metric by a general metric $\,\eta_{\mu\nu}\to g_{\mu\nu}$\,
(together with the rescaling of integral measure $\,\di^4x\to \di^4x\,\sqrt{-g}$\,).
The other is to replace the partial derivative by covariant derivative
$\,\pd_\mu^{}\to\nabla_\mu^{}$,\, where $\,\nabla_\mu$\, is adapted for $g_{\mu\nu}$
with  $\,\nabla_\lambda^{} g_{\mu\nu}^{}=0\,$.\,
For fermionic fields, the vierbein and spin connection are introduced.
With these, we could conclude that the joint action
$\,S_{\text{EH}}^{}+S_{\text{G}4}^{}+S_{\text{MC}}^{}$\,
provides an effective description of the SM with gravitation.
But, this action is incomplete up to dimension-4 operators.
There is a unique dimension-4 operator which couples the Higgs doublet $H$
to the scalar curvature $R$,\,
and thus should be added to the above action,
\beqa
\label{eq:NMC}
S_{\text{NMC}} \,=\, \int\!\!\di^4x\sqrt{-g\,}\,\xi\mathcal{R}H^\dag H \,,
\eeqa
where $\,\xi\,$ is a dimensionless coupling.
This term is conventionally called nonminimal coupling term
since it does not follow the rule of minimal coupling.
We note that up to dimension-4 operators, the nonminimal coupling could only appear
for the spin-0 field. This fact adds a unique feature to the Higgs boson in the SM.

Hence, the complete Lagrangian up to dimension-4 operators and joining both
the SM and GR should take the following form,
\beqa
\label{Action}
S ~=~ S_{\text{EH}}^{} + S_{\text{G}4}^{} + S_{\text{MC}}^{} + S_{\text{NMC}}^{} \,.
\eeqa
The non-minimal coupling term $\,S_{\text{NMC}}^{}\,$
is generally covariant and respects all known symmetries of the SM.
We further note that $\,\xi\to 0\,$ does not enlarge the symmetry,
and a nonzero $\,\xi\,$ will still be generated by loop diagrams even if one sets
$\,\xi=0\,$ classically \cite{RGxi}.
Another special value is the conformal coupling $\,\xi=-1/6$,\,
which makes the theory Weyl-invariant for massless scalar field.
But, the SM Higgs doublet is not massless, and the Weyl symmetry
(which reduces to the conformal symmetry in flat spacetime) is also not a symmetry of the SM.
All these facts imply that the size of this dimensionless nonminimal coupling $\,\xi\,$
can be rather large \emph{a priori}. In fact, a large $\,\xi\,$ around the order of
$\,10^4\,$ has been put in use for the Higgs inflation models
\cite{bezrukov,Barvinsky:2008ia,DeSimone:2008ei,Barvinsky:2009fy,Barvinsky:2009ii,BgdDep2010,BgdDep}
in which the Higgs boson is responsible for two distinctive physical processes,
namely, it drives inflation at a typically very high inflation scale, and triggers
the electroweak symmetry breaking at weak scale.\footnote{
Ref.\,\cite{Barvinsky:2008ia} first observed that the value of Higgs boson mass
can be directly related to the CMB parameters in the Higgs inflation.}

Recently, Ref.\,\cite{atkins} derived an interesting bound on $\xi$ from the LHC Higgs data,
by assuming the 125\,GeV boson to be the SM Higgs boson.
As will be shown in Sec.\,\ref{formal}, a large $\,\xi$\, will cause a universal suppression
of the Higgs boson coupling with all other SM particles.
Thus, using the measured Higgs signal strengths in 2012, Ref.\,\cite{atkins} derived an upper bound,
$\,|\xi|<2.6\times 10^{15}$.\,
Furthermore, in the recent study \cite{XRH}, we derived the perturbative unitarity bound on $\,\xi\,$
by analyzing the coupled-channel longitudinal weak boson scattering under flat spacetime background,
and we demonstrated the longitudinal-Goldstone boson equivalence theorem\,\cite{He:1997zm}
in the presence of $\,\xi\,$ coupling.
In \cite{XRH}, we did calculations only in Einstein frame,
where the nonminimal coupling is transformed away by redefining the metric tensor.
But, we may also perform the analysis in Jordan frame,
i.e., with the field variables written in the action (\ref{Action}).
The physical (in)equivalence between Jordan frame and Einstein frame is a subtle issue
which still lacks a full consensus (especially at the quantum level)
\cite{conformal,Bezrukov:2009db,Barvinsky:2009fy,Barvinsky:2009ii,JE-new}.
Then, it is desirable to perform an independent analysis within the Jordan frame,
in comparison with our previous Einstein frame analysis \cite{XRH}.
In this paper, we will show that the same results can be inferred from Jordan frame
with a fully different set of Feynman rules (Appendix\,\ref{A:FR}).
This serves as a valuable consistency check of our analysis.
For the first time, we will further explicitly prove
the longitudinal-Goldstone equivalence theorem with nonzero $\,\xi\,$ in Jordan frame.
It provides another nontrivial consistency check
on our Jordan-frame analysis.

There are some discussions on the unitarity issue with the nonminimal coupling for the purpose
of Higgs inflation models \cite{ext}.  People usually applied power-counting arguments to get
the scalar or vector boson scattering amplitudes, and estimate the allowed regime
for the perturbative calculations of Higgs inflation.
At the first sight, the unitarity bound is around $\,\Mp/\xi$\, for $\,\xi\gg 1\,$,\,
which is lower than the typical inflation scale $\,\Mp/\sqrt{\xi}\,$.\,
Later studies \cite{BgdDep2010,BgdDep} suggested that unitarity bounds vary with respect to classical background
of inflaton field. Considering this background dependence, the unitarity bound is expected to be relaxed
to $\,O(\Mp/\!\sqrt\xi)$\, in the inflationary era.
This implies that the perturbative analysis of inflation dynamics and primordial fluctuation
would be reliable.  These discussions are suggestive, but are only qualitative power-counting estimates
with different focus and context. In this paper, we will systematically extend
our analysis to quantitatively derive the unitarity constraints for a generical
Higgs inflaton background. We will consider the conventional Higgs inflation\,\cite{bezrukov}
and the improved models\,\cite{B2new1,B2new2,B2new3}
in light of the recent BICEP2 data \cite{BICHEP2}.

This paper is organized as follows.
In Sec.\,\ref{formal}, we will outline the formulation with nonminimal coupling
for the Jordan and Einstein frames.
In Sec.\,\ref{WBscat}, we study the longitudinal weak boson scatterings in both frames.
This gives a systematical extension of our previous short study\,\cite{XRH}
which was for the Einstein frame alone
and at the lowest order of $\,1/\Mp^{2}\,$.\,
In Sec.\,\ref{sec3.3}, we quantitatively derive the perturbative unitarity
bound on the Higgs-curvature coupling $\,\xi\,$ in both the electroweak vacuum and the large field background.
We further study $\xi$-dependent weak boson scattering cross sections
at the scales of $\order{1-30}$TeV energies,
which serve as the new probes of $\,\xi\,$ at the upcoming LHC\,(14\,TeV) and the future high energy
$pp$ colliders ($50-100$\,TeV) \cite{FCC}.
In Sec.\,\ref{BkgD}, we systematically extend our analysis to
the Higgs inflation models in the large field background.
We conclude in Sec.\,\ref{conclusion}.
Finally, we present the necessary Feynman rules
for both Jordan and Einstein frames in Appendix\,\ref{A:FR}.

\vspace*{2mm}
\section{Higgs-Gravity Interactions in Jordan and Einstein Frames}
\label{formal}
\vspace*{2mm}

In this section, we discuss the formulation for the nonminimal gravitational coupling of the SM Higgs doublet,
which will serve as the base for our systematical studies in the following sections\,3--4.
The set of field variables that manifests this nonminimal coupling \eqref{eq:NMC} is conventionally
called Jordan frame, which is part of the joint effective action (\ref{Action}).
One can make a proper Weyl transformation on the metric tensor, under which the
the nonminimal coupling term \eqref{eq:NMC} is fully transformed away.
This new metric is usually called Einstein frame, in which the original Einstein equation of GR holds.
For clarity we will use a superscript $(J)$ to label the Jordan frame metric and other geometric quantities,
while the Einstein frame quantities are presented without superscript so long as they appear confusion-free.

\vspace*{2mm}
\subsection{Higgs-Gravity Interactions in Jordan Frame}
\vspace*{2mm}

 By definition, the action in Jordan frame takes the form of (\ref{Action}).
 Hence, the nonminimal coupling term $\,\xi\mathcal{R}^{(J)}H^\dag H$\, is manifest in Jordan frame.
 To analyze the gravitational interactions of Higgs boson and weak gauge bosons,
 we extract from (\ref{Action}) the terms with scalar-curvature, weak gauge bosons and Higgs doublet,
 and organize them in a more transparent way,
\begin{align}
  \label{Action_Jordan}
  \hspace*{-3mm}
  S_{\text{J}} ~=\, & \int\!\!\di^4x\,
  \mbox{$\sqrt{-g^{(J)}}$} \Big[\(\fr{1}{2}M^2+\xi H^\dag H\)\mathcal{R}^{(J)}
   -\sum_j\!\fr{1}{4}F_{\mu\nu j}^aF^{\mu\nu a}_j+(\D_\mu H)^\dag(\D^\mu H) - V(H)\Big] ,~~~
\end{align}
where $\mathcal{R}^{(J)}$ is the Ricci scalar associated with the Jordan frame metric
$\,g_{\mu\nu}^{(J)}\,$,\,  and $\,F_{\mu\nu i}^a=(W_{\mu\nu}^a,B_{\mu\nu})$
are gauge field strengths of the electroweak gauge group $SU(2)_L^{}\otimes U(1)_Y^{}$.\,
In this action we do not display the cosmological constant term
as well as the dimension-4 operators of pure gravity,
since they are irrelevant to the current study.
The Higgs doublet $H$ is parameterized as,
$\,H=\big(\pi^+,\,\fr{1}{\sqrt{2}\,}(v + \hat\phi + \ii \pi^0)\big)^T$\,
and has the potential
$\,V(H)=\lam\(H^\dag H-\fr{1}{2}v^2\)^2$.\,
The potential reaches its minimum when Higgs doublet acquires a nonzero vacuum expectation value (VEV),
$\,\la H\ra =\fr{1}{\sqrt 2}(0,v)^T$.\,
Substituting this VEV into the action (\ref{Action_Jordan}),
we see that the nonminimal coupling term gives a contribution to
the pure gravitational term $\,\sqrt{-g^{(J)}}\mathcal{R}^{(J)}$,\,
and leads to an effective Einstein-Hilbert action,
\beqa
\int\!\!\di^4x\mbox{$\sqrt{-g^{(J)}}$}\Big(\fr{1}{2}M^2+\xi H^\dag H\Big)\mathcal{R}^{(J)}
~\supset~
\int\!\!\di^4x\mbox{$\sqrt{-g^{(J)}}$}\,\fr{1}{2}\!\(M^2+\xi v^2\)\mathcal{R}^{(J)} \,.
\eeqa
Thus, we identify the coefficient, $\,M^2+\xi v^2=\Mp^2$,\,
to reproduce the correct Newton constant.

In Jordan frame, gravitons make important contributions to the weak gauge boson scattering amplitudes.
Hence, we perturb the metric, $\,g_{\mu\nu}^{(J)}=\bar g_{\mu\nu}^{(J)}+\ka \hat h_{\mu\nu}$\,,\,
with the background value $\,\bar g_{\mu\nu}^{(J)}$\, identified as the flat Minkowski metric,
$\,\bar g_{\mu\nu}^{(J)}=\eta_{\mu\nu}^{}$.\,
The perturbation coupling $\,\ka=\sqrt{2}\Mp^{-1}$\, is chosen such that the spin-2 component
in $\,\hat h_{\mu\nu}$\, has the correct normalization.
A special feature of the Jordan frame is that both the perturbed metric $\,\hat h_{\mu\nu}^{}\,$
and the Higgs field $\,\hat \phi\,$ are not canonically normalized due to kinetic mixings.
To see this, we inspect the kinetic terms for $\,\hat h_{\mu\nu}^{}\,$ and $\,\hat \phi\,$
from (\ref{Action_Jordan}),
\begin{align}
\label{eq:S-J}
\hspace*{-3mm}
  S_{\text{J}}^{} \,\supset\,
  \fr{1}{4}\big(\hat h\square\hat h-\hat h^{\mu\nu}\square\hat h_{\mu\nu}+2\hat h_{\mu\lam}\pd^\mu\pd_\nu
  \hat h^{\nu\lam}-2\hat h\pd^\mu\pd^\nu \hat h_{\mu\nu}\big) -\fr{1}{2}\hat\phi\square\hat\phi
  +\xi\ka v\hat\phi\big(\square\hat h-\pd^\mu\pd^\nu \hat h_{\mu\nu}\big),~~
\end{align}
where $\,\hat h\equiv \hat h_\mu^\mu$\,.\,
Thus, we define another set of canonically normalized fields $\,h_{\mu\nu}^{}\,$ and $\,\phi\,$,
\beqa
\label{eq:kineticnorm}
\hat h_{\mu\nu}^{} = h_{\mu\nu}^{}-\eta_{\mu\nu}^{}\xi\ka v\zeta \phi \,,
\qquad
\hat\phi=\zeta \phi \,,
\qquad \qquad
\eeqa
with the rescaling factor,
\begin{eqnarray}
\label{rescalezeta}
\zeta \,\equiv\,
\(1+3\xi^2\ka^2v^2\)^{-\frac{1}{2}} \,=\, \(1+6\xi^2v^2/\Mp^2\)^{-\frac{1}{2}} \,.
\end{eqnarray}
After this field rescaling, we readily verify that both new variables
$\,h_{\mu\nu}^{}$\, and $\,\phi\,$ are canonically normalized, with the correct kinetic term,
\beqa
\label{eq:L-kin}
  \ld_{\text{kin}}^{} \,=\, \fr{1}{4}\big(h\square h- h^{\mu\nu}\square h_{\mu\nu}+2h_{\mu\lam}\pd^\mu\pd_\nu h^{\nu\lam}-2h\pd^\mu\pd^\nu h_{\mu\nu}\big)-\fr{1}{2}\phi\square\phi \,.
\eeqa
The above field redefinitions have two general consequences.
First, from the rescaling of Higgs field $\,\hat\phi=\zeta\phi$\,,\,
we see that all SM couplings of gauge fields and fermions to the Higgs boson $\,\phi\,$
receive a universal rescaling factor $\,\zeta$\, associated with each Higgs field in the vertex.
This rescaling factor $\,\zeta < 1\,$
causes a universal suppression on the Higgs boson couplings with all other SM particles.
Second, the original perturbed metric $\,\hat h_{\mu\nu}^{}\,$ contains a portion of
canonical Higgs field $\,\phi$\,.\,
Hence, there will be new Higgs couplings to other SM particles which are induced from gravitational interactions.
Since the graviton $\,\hat h_{\mu\nu}^{}$\, couples to matter fields through their stress tensor
(with the exception of nonminimal coupling to Higgs doublet), the graviton-induced Higgs couplings to SM particles
are all suppressed by the small parameter $\,\ka$\,.\,
But, as we will show, the gravity contributions can be significant
when the nonminimal coupling $\,\xi$\, is large.

To compute scattering amplitudes with graviton exchange,
we fix the gauge symmetry of linearized general covariance
by choosing the harmonic gauge. This amounts to adding the
gauge-fixing term,
\beqa
\ld_{\text{GF}}^{} \,=\, \frac{1}{2\al}(\pd^\lam h_{\lam\mu}-\fr{1}{2}\pd_\mu h)^2 ,
\eeqa
with $\,\al\,$ as the gauge parameter.
Then, the kinetic term of the graviton can be inverted to give the graviton propagator,
\beqa
  G_{\mu\nu,\rh\si}^{(h)}(p) \,=\FR{\ii}{\,p^2+\ii\ep\,}
  \[2\eta_{\mu(\rh}^{}\eta_{\si)\nu}^{} - \eta_{\mu\nu}^{}\eta_{\rh\si}^{}
        -(1\!-\!\al)\FR{4p_{(\mu}^{}\eta_{\nu)(\rh}^{}p_{\si)}^{}}{p^2}\].
\eeqa
The gauge-invariance of the scattering amplitudes can be checked by showing
that the $\al$-dependent pieces vanish.
This is necessarily true if the graviton is coupled to external on-shell particles,
which are subjected to the energy-momentum conservation $\,\pd^\mu T_{\mu\nu}^{}=0$\,
in the flat spacetime. We note that the nonminimal coupling term will spoil the covariant conservation of
$\,T_{\mu\nu}^{}\,$  with curved background, but it does not affect the flat-space conservation law,
$\,\pd^\mu T_{\mu\nu}^{}=0$,\, since the Ricci scalar vanishes in this case.

The relevant new interactions involving canonically normalized graviton $\,h_{\mu\nu}^{}\,$
and Higgs $\,\phi\,$ can be extracted from the action (\ref{Action_Jordan}),
\begin{align}
  \label{Int_Jordan}
  \Delta\ld_{\text{int}}^{h\phi} \,=\,
  \kappa\(h_{\mu\nu}^{} - \eta_{\mu\nu}^{}\xi\ka v\zeta\phi\)T^{\mu\nu}_{},
\end{align}
where $\,T^{\mu\nu}\,$ is the energy-momentum tensor of matter fields,
including gauge bosons and Higgs doublet for the current study.
In the Feynman-'t\,Hooft gauge, we have
\begin{align}
  T^{\mu\nu}
  \,= &~\sum_j^{}\Big(F^{\mu\lam a}_j {F_{\lam}}_j^{\nu a}
      +\fr{1}{4}g^{\mu\nu}F_{\rh\si j}^aF^{\rh\si a}_j\Big)
      + m_W^2\Big(W^{\mu a}W^{\nu a}-\fr{1}{2}g^{\mu\nu}W_\lam^a W^{\lam a}\Big)
\n\\
   &~+\pd^\mu\pi^a\pd^\nu\pi^a+\pd^\mu\phi\pd^\nu\phi
    -\fr{1}{2}g^{\mu\nu}
    \Big((\pd_\lam^{}\pi^a)^2+(\pd_\lam^{}\phi)^2 - m_W^2\pi^a\pi^a - m_\phi^2\phi^2\Big),
\end{align}
where we only keep contributions from kinetic terms and mass terms for the current purpose.
We have also set the weak mixing angle $\,\theta_W^{}=0\,$ for simplicity,
and dropped terms from the gauge-fixing and Faddeev-Popov ghosts
which are irrelevant to the tree-level processes in the present work.

\vspace*{2mm}
\subsection{Higgs-Gravity Interactions in Einstein Frame}
\vspace*{2mm}

The Einstein frame is realized by the field redefinition,
$\,g_{\mu\nu}^{(E)}=\Omega^2 g_{\mu\nu}^{(J)}$, with
\begin{eqnarray}
\label{Omega}
\Omega^2 ~=\, \frac{\,M^2\!+2\xi H^\dag H\,}{\Mp^2} \,.
\end{eqnarray}
We see that this field redefinition takes the form of Weyl transformation,
and the Weyl factor $\,\Omega^2\,$ depends on spacetime coordinates through
its dependence on the Higgs-doublet-bilinear term $\,H^\dag H$.\,
This transformation is also known as ``conformal transformation''
in the literature of general relativity.
(Here we avoid the use of this terminology since it is also used for a special type of
spacetime coordinate transformations that leave the metric invariant up to a spacetime-dependent factor.)
We note that the field redefinition here is not generated by a spacetime coordinate transformation.
Hence, one would not expect that the physics described by these two frames to be equivalent \emph{a priori,}
as extensively discussed in the literature 
\cite{Barvinsky:2009fy,Barvinsky:2009ii,conformal,Bezrukov:2009db,JE-new}.
Thus, in a given study, one should be careful to check whether the two frames describe the same physics or
which frame has a better use.
But, this ambiguity is absent for our current study of gauge boson scattering processes.
As will be explained below, one can always set the background spacetime metric
to be flat to a good approximation,
and the two ``frames'' just correspond to different choices of field variables and
thus should yield the same physical results.

Let us convert the Jordan frame action (\ref{Action_Jordan}) into the Einstein frame.
To achieve this, we note that the Weyl transformation of Ricci scalar takes the form,
\begin{align}
  \mathcal{R}^{(J)} =\,
  \Omega^2\Big[\mathcal{R}-6g^{\mu\nu}_{}\nabla_\mu^{}\nabla_\nu^{}\log\Omega
  +6g^{\mu\nu}_{}\big(\nabla_\mu^{}\log\Omega\big)\big(\nabla_\nu^{}\log\Omega\big)\Big],
\hspace*{8mm}
\end{align}
where the unlabeled quantities on the right-hand side,
such as Ricci scalar $\,\mathcal{R}$\,,\, the covariant derivative $\nabla_\mu^{}$\,,\,
are associated with the Einstein frame metric $\,g_{\mu\nu}^{}=g_{\mu\nu}^{(E)}$.\,
We drop the superscript $(E)$ for these quantities to simplify the notation.
Substituting this transformation into (\ref{Action_Jordan}),
we obtain the Einstein frame action,
\begin{align}
  \label{Action_Einstein}
  S_{\text{E}}^{} \,=\int\!\!\di^4x\,\sqrt{-g\,}
  \bigg[ & \fr{1}{2}\Mp^2 \mathcal{R}-\sum_j^{}\fr{1}{4}F_{\mu\nu j}^aF^{\mu\nu a}_j
          -\FR{3\xi}{\Omega^2}\nabla^2(H^\dag H)+\FR{9\xi^2}{\Mp^2\Omega^4}\big(\nabla_\mu^{}(H^\dag H)\big)^2
  \n\\
  & +\FR{1}{\Omega^2}(\D_\mu H)^\dag(\D^\mu H)-\FR{1}{\Omega^4}V(H)\bigg].
\end{align}
 In comparison with the Jordan frame action (\ref{Action_Jordan}),
 we see that the nonminimal coupling term $\,\xi\mathcal{R}^{(J)}H^\dag H$\,
 is fully transformed away.
 In consequence, the equation of motion (from varying the metric) reproduces
 the conventional Einstein equation, and the graviton has the correct kinetic term as in the GR.
 Under the linearized expansion $\,g_{\mu\nu}^{}=\bar{g}_{\mu\nu}^{}+\kappa h_{\mu\nu}^{}$,\,
 we can write the Weyl transformation $\,g_{\mu\nu}^{}=\Omega^2 g_{\mu\nu}^{(J)}$\,
 in the perturbative form,
\begin{eqnarray}
\label{eq:EFfieldRedef}
&& \bar{g}_{\mu\nu}^{}+\kappa h_{\mu\nu}^{}
   \,= \(\!1+\frac{\xi}{M_p^2}2v\hat\phi+...\)\!\(\eta_{\mu\nu}^{}+\kappa\hat{h}_{\mu\nu}^{}\),
\nonumber\\[2mm]
&& \bar{g}_{\mu\nu}^{} = \eta_{\mu\nu}^{},  \quad
h_{\mu\nu}^{} = \hat{h}_{\mu\nu}^{} + \eta_{\mu\nu}^{}\xi\kappa v\hat{\phi} \,,
\end{eqnarray}
where the background in Einstein frame is flat as well.
Since $\,\Omega^2\,$ contains the Higgs-doublet-bilinear term $\,H^\dag H$,\,
new interactions with Higgs fields are induced,
where the Higgs field $\,\hat\phi\,$ is not normalized canonically.
With the flat spacetime background
$\,\bar{g}_{\mu\nu}^{}=\eta_{\mu\nu}^{}$\, and substituting
$\,H=\big(\pi^+\!,\,\fr{1}{\sqrt{2}\,}(v + \hat\phi + \ii\pi^0)\big)^T$\,
into the action (\ref{Action_Einstein}), we derive the following kinetic term
(up to bilinear fields) for the Higgs and Goldstone bosons,
\begin{align}
  \ld_{\text{kin}}^{} \,=\, \FR{1}{2}\bigg(\!1+\FR{6\xi^2v^2}{\Mp^2}\!\bigg)
  (\pd_\mu \hat\phi)^2+\pd_\mu\pi^-\pd^\mu\pi^++\FR{1}{2}(\pd_\mu\pi^0)^2.
\end{align}
As is clear, to normalize the Higgs kinetic term requires a field redefinition,
$\,\hat\phi=\zeta\phi$\,,\,
where the rescaling factor $\,\zeta=\(1+6\xi^2v^2/\Mp^2\)^{-1/2}$\,
takes the same form as in the Jordan frame.
As compared to \eqref{eq:kineticnorm}-\eqref{eq:L-kin} in the Jordan frame,
we see that the canonical Higgs fields in both frames are defined in the same way up to quadratic terms.
Thus, we can directly infer the tree-level Higgs mass,
$\,m_\phi^2=2\lam v^2\zeta^2$,\,
and all the Higgs couplings in the SM under the rescaling for each Higgs field
($\,\hat\phi=\zeta\phi$\,).\,
For the Higgs self-couplings, we have further contributions from higher dimensional operators
besides the rescaling of $\,\zeta$\,.\, These new couplings contain derivatives,
so we expect them to become important for scattering processes at high energies.
This means that for the present study of the high energy scattering of weak gauge bosons,
we should take all these $\xi$-corrections into account.
To the $\,\order{\Mp^{-2}}$\,,\, we extract these new interactions from
the action (\ref{Action_Einstein}),
\begin{align}
  \label{DL_int_ss}
  \Delta\ld_{\text{int}}^{ss}
  \,=\,
   & -\FR{\xi}{2\Mp^2}\Big(|\pd_\mu\pi|^2+\zeta^2(\pd_\mu \phi)^2\Big)\Big(|\pi|^2+\zeta^2\phi^2+2v\zeta \phi\Big)
\n\\[1.5mm]
   &-\FR{3\xi^2}{4\Mp^2}\Big(|\pi|^2+ \zeta^2\phi^2+2v\zeta \phi\Big)\pd^2\Big(|\pi|^2+\zeta^2 \phi^2+2v\zeta \phi\Big),
\hspace*{10mm}
\end{align}
where $\,|\pi|^2=2\pi^+\pi^-\!+(\pi^0)^2$\, and
$\,|\pd_\mu\pi|^2=2\pd_\mu\pi^+\pd^\mu\pi^-\!+(\pd_\mu\pi^0)^2$\,.\,
For the Higgs couplings with weak gauge bosons, we infer,
\begin{align}
\label{DL-int_pg}
  \Delta\ld_{\text{int}}^{\phi g}
  \,=\, \(2m_W^2 W_\mu^+W^{\mu -} \!+m_Z^2 Z_\mu^2\)\!
  \left[\(\! 1-\FR{\xi v^2}{\Mp^2}\)\!\frac{\zeta}{v}\, \phi
  + \!\(\! 1-\FR{5\xi v^2}{\Mp^2}\)\!\FR{\zeta^2}{2v^2}\,\phi^2\right] \!.
\hspace*{5mm}
\end{align}

For completeness, we further inspect the fermion sector under the present formalism.
For Dirac fermion $\Psi$ in the curved background, we can write its kinetic term,
\beqa
S_{\text{f}}^{} \,=
\int\!\!\di^4x\,\det(e_\nu^q)\bar\Psi\ga^p e^\mu_p
\Big(\ii\pd_\mu-\FR{1}{2}\omega_\mu{}^{mn}\si_{mn}\Big)\Psi \,,
\eeqa
where $\,e_\nu^q\,$ and $\,\omega_\mu{}^{mn}\,$ are vierbein and spin-connection,
and $\,\si_{mn}=\fr{\ii}{2}[\ga_m,\ga_n]$\,.\,
We identify the background to be the Jordan frame,
which is related to the flat Einstein frame by
$\,g^{(J)}_{\mu\nu}=\Omega^{-2}\eta_{\mu\nu}^{}$\,.\,
Then, we can derive expressions for the vierbein and spin-connection
in terms of Weyl factor $\,\Omega$\,,
\beqa
  e_\mu^m=\Omega^{-1}\de_\mu^m \,,
\hspace*{10mm}
  \omega_\mu{}^{mn}=-\Omega^{-1}\(\de_\mu^m\pd^n\Omega-\de_\mu^n\pd^m\Omega\) ,
\eeqa
and the above kinetic term becomes
\beqa
\label{Action_f}
S_{\text{f}}^{} \,=\,
\ii\int\!\!\di^4x\,\Big(\fr{1}{\Omega^3}\bar\Psi\sla\pd\Psi
+\fr{3}{\Omega^4}\bar\Psi(\sla\pd\Omega)\Psi\Big) \,.
\eeqa
This shows that the fermion sector also receives corrections from nonminimal coupling $\xi$
through the Weyl factor $\,\Omega\,$.\,
When expanded in terms of $\Mp^{-1}$, these corrections correspond to higher dimensional operators.
At the $\,\order{\Mp^{-2}}$,\, the coefficients of these operators could have linear $\xi$-dependence
at most. Hence, in the large $\xi$ regime, such corrections from the fermion sector are
relatively small as compared with those in the gauge sector.

In summary, we discuss in this section the formulation of the SM coupled to gravity through
both minimal and nonminimal couplings in the electroweak vacuum. We derive the relevant Lagrangians
and Feynman rules in terms of canonically normalized fields for both Jordan and Einstein frames,
which are shown in Appendix\,\ref{A:FR}.
For the leading order $\xi$ corrections, the graviton contributions are important in Jordan frame,
while in Einstein frame it is the effective higher dimensional interactions that are really relevant.
From these analyses, we see the advantage of Einstein frame, in which the complicated graviton contributions
are transformed away and result in the higher dimensional effective Higgs operators.
The latter makes the $\xi$-induced modifications to the SM sector transparent, and allow us
to perform the calculations in a much easier manner.

\vspace*{2mm}
\section{Weak Boson Scattering in Jordan and Einstein Frames}
\label{WBscat}

The action (\ref{Action}) retains the leading terms
in the low energy effective theory that combines the SM and GR.
They are perturbatively non-renormalizable in the conventional sense.
In both Jordan and Einstein frames, this fact will certainly manifest
through the $\xi$-dependent effective operators with Higgs doublets.
In consequence, we observe that for the high energy scattering of longitudinal weak bosons,
the scattering amplitudes will generally exhibit non-canceled $E^2$ behaviors
(with $E$ the scattering energy).
For $\,|\xi|\gg 1\,$,\, the non-canceled $E^2$ terms induced by the nonminimal coupling $\,\xi\,$
will dominate the amplitudes, and lead to violation of perturbative unitarity.
Based on the longitudinal-Goldstone boson equivalence theorem (ET) \cite{He:1997zm},
we expect that the same non-canceled $E^2$ behavior can be derived from the corresponding
Goldstone boson scattering amplitudes. In this section, we will systematically analyze
the weak boson scattering with nonzero $\,\xi\,$ and demonstrate the ET in both
Jordan and Einstein frames. Such demonstrations are highly nontrivial.
In Einstein frame, we note that the $\xi$-dependent scalar derivative interactions (\ref{DL_int_ss})
take very different forms from the new Higgs-gauge boson couplings (\ref{DL-int_pg}).
In Jordan frame, we have leading contributions from graviton exchanges.
Furthermore, computing the longitudinal and Goldstone amplitudes in both frames
will help to demonstrate the equivalence between the two frames, and
serve as valuable consistency checks for our analyses.

Then, we derive the perturbative unitarity bound on the nonminimal coupling via coupled channel analysis.
For this, we will study all neutral channels of weak bosons
and Higgs boson scattering with normalized initial/final states,
$\,|W_L^+W_L^-\ra$, $\fr{1}{\sqrt2}|Z_L^0Z_L^0\ra$, $\fr{1}{\sqrt2}|\phi\phi\ra$, and $|Z_L^0\phi\ra$,\,
as well as the corresponding Goldstone boson processes with initial/final states
$\,|\pi^+\pi^-\ra$, $\fr{1}{\sqrt 2}|\pi^0\pi^0\ra$, $\fr{1}{\sqrt 2}|\phi\phi\ra$,\,
and $|\pi^0\phi\ra$.\, As we will show, the $\,2\to 2\,$ process involving four identical
external particles has vanishing $E^2$ term from non-minimal coupling,
due to the crossing symmetry among $(s,\,t,\,u)$ channels.
The rest of non-trivial processes can be classified into two categories,
depending on whether the in/out states contain Higgs boson or not.
The amplitudes in each category have similar form.
In the following, we will investigate these two types of processes
for the weak boson and Goldstone boson scatterings in each frame.
We explicitly demonstrate the ET in both frames and show the equivalence of computations
between the two frames.
Finally, we will derive the perturbative unitarity bound on the Higgs-gravity coupling $\,\xi\,$
and analyze the probe of $\,\xi$-induced weak boson scatterings
at the LHC\,(14\,TeV) and the future high energy $pp$ colliders ($50-100$\,TeV) \cite{FCC}.
These also systematically extend our recent short study\,\cite{XRH}
(which was for the Einstein frame alone and only to the first order of $\,1/\Mp^{2}\,$).

\vspace*{2mm}
\subsection{Analysis in Jordan Frame}
\label{JFanalysis}

\begin{figure}[h]
\begin{center}
\vspace*{4mm}
  \includegraphics[width=0.92\textwidth]{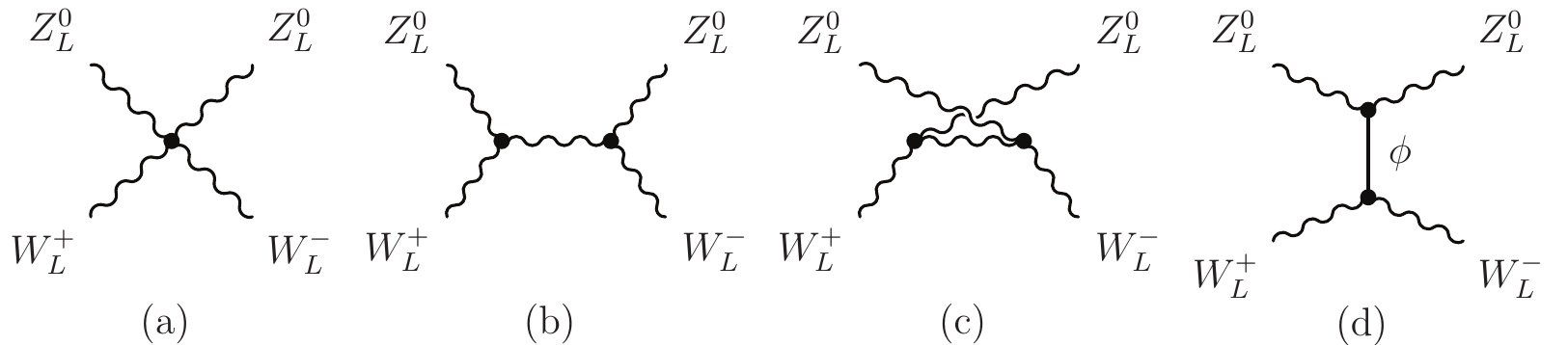}
\caption{Weak boson scattering process $\,W_L^+W_L^- \to Z_L^0Z_L^0\,$ at tree-level, via SM-type
 diagrams. The nonminimal coupling $\,\xi\,$ leads to anomalous Higgs-gauge-boson couplings in (d).}
  \label{fig:1}
\end{center}
\end{figure}

We start our analysis in Jordan frame.
The first category of processes is for the weak boson scatterings,
with initial/final states containing no Higgs boson $\,\phi$\,.\,
For demonstration, we first consider the sample process $\,W_L^+W_L^-\to Z_L^0Z_L^0$\,.\,
There are four SM-types of Feynman diagrams contributing to this process at tree-level,
as shown in Fig.\,\ref{fig:1}, except that the invoked Higgs-gauge-boson couplings
are modified from the corresponding SM couplings.
We present the relevant Feynman rules in Appendix\,\ref{A:FRJ}.

While the pure gauge couplings remain unchanged in Jordan frame,
there are two new ingredients in Higgs couplings.
One is the Higgs rescaling factor $\,\zeta\,$ [cf.\ \eqref{eq:kineticnorm}-\eqref{rescalezeta}]
for each vertex with Higgs field(s),
and the other is the Higgs component in the original perturbed metric $\,\hat{h}_{\mu\nu}^{}$
[cf.\ \eqref{eq:kineticnorm},\eqref{Int_Jordan}].\,
The former causes $\,\order{\xi^2/\Mp^2}\,$ corrections to the SM Higgs coupling,
while the latter induces a new Higgs coupling of $\,\order{\xi/\Mp^2}$\,
to the trace of stress tensor of gauge bosons.
With these, we compute the diagram with $s$-channel Higgs exchange in Fig.\,\ref{fig:1}(d),
\begin{align}
  \label{T_WWZZ4_JF}
  \mathcal{T}_{\phi}^s (W^+_LW^-_L \!\to Z_L^0Z_L^0)
  ~=\, & -\!\bigg(1-\FR{\xi v^2}{\Mp^2}\bigg)^{\!2} \FR{\zeta^2 E^4}{\,v^2(E^2-m_\phi^2)\,},
\end{align}
where $\,E\,$ is the center-of-mass energy.
At $\order{E^2}$, the other three pure gauge-boson diagrams in Fig.\,\ref{fig:1}
contribute to the amplitude with $\,\order{E^2/v^2}$\, terms.
Thus, we deduce the full amplitude to all orders in $\,\xi^2/\Mp^{2}$\,,
\beq
\label{TWWZZ_Jor_Lon}
\begin{aligned}
\T(W_L^+W_L^- \!\to Z_L^0Z_L^0)
~=~& \bigg[1-\Big(1-\FR{\xi v^2}{\Mp^2}\Big)^2\zeta^2\bigg]\FR{E^2}{v^2} + \order{E^0}
\\[2mm]
~=~& \bigg[\FR{\,6\xi^2\!+ 2\xi\,}{\Mp^2}-\FR{36\xi^4v^2}{\Mp^4}\bigg]E^2
     + \order{E^0, \Mp^{-6}} \,, \hspace*{12mm}
\end{aligned}
\eeq
where in the second step we have expanded the amplitude up to $\,\order{\xi^4/\Mp^4}\,$
for comparison.
We have also kept the first subleading term of $\,\order{\xi/\Mp^2}\,$
at the lowest order in $\,1/\Mp^2\,$ expansion.
As will be clear later, for $\,|\xi|\gg \order{1}\,$,\, the amplitudes are dominated
by the leading terms with the form of $\,\(\xi/\Mp^{}\)^{2n}$\,.

\begin{figure}[h]
\begin{center}
\includegraphics[width=0.5\textwidth]{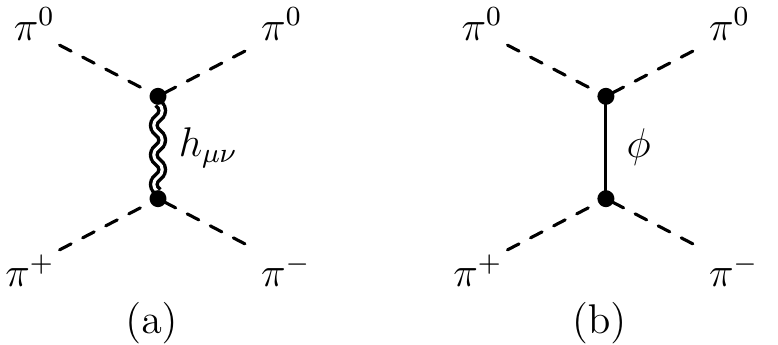}
\vspace*{-2mm}
\caption{Goldstone boson scattering process $\,\pi^+\pi^-\to \pi^0\pi^0$\, in Jordan frame,
via the $s$-channel graviton exchange in plot-(a), and the $s$-channel Higgs boson exchange in plot-(b).}
\label{fig:2}
\end{center}
\end{figure}

Next, we compute the amplitude of the corresponding Goldstone boson scattering
$\,\pi^+\pi^-\to \pi^0\pi^0$\,.\,
At the order of $\,E^2/\Mp^2\,$,\,
the only relevant Feynman diagram comes from the $s$-channel graviton exchange,
as shown in Fig.\,\ref{fig:2}(a).
Thus, we deduce
\begin{align}
\label{eq:WWZZgoldstone}
  \T_h^s(\pi^+\pi^- \!\to \pi^0\pi^0) ~=
  \left[\(6\xi^2\!+2\xi\) + \FR{\,1\!-\cos^2\theta\,}{4}\right]\!\FR{E^2}{\,\Mp^2\,}
  + \order{E^0} \,,
\end{align}
where $\,\theta\,$ is the scattering angle.
Eq.\,\eqref{eq:WWZZgoldstone} is dominated by the $\xi$-dependent contributions for $\,|\xi| \gg 1\,$.\,
There is no correction of $\,\order{1/\Mp^4}$\, or higher to this tree-level process.
On the other hand, for the diagram with the $s$-channel Higgs exchange in
Fig.\,\ref{fig:2}(b), its contribution to the $E^2$-amplitude starts only at
the order of $\,\Mp^{-4}$\,.\,  At each order in $\Mp^{-1}$, we only consider leading
$\xi$-terms of the form $\,\(\xi/\Mp^{}\)^{2n}$,\,  except that at the lowest order
of $\,\Mp^{-2}$\,,\, we retain the subleading term of $\,\order{\xi/\Mp^{2}}$.\,
It is clear that the leading terms of $\,\(\xi/\Mp^{}\)^{2n}$\, always dominate the amplitudes
for $\,|\xi|\gg 1\,$.\, We note that such leading terms arise from
the $\xi$-enhanced effective operators and the rescaling factor $\,\zeta$\, (cf.\ Sec.\,2).
Thus, we infer the following expression for Fig.\,\ref{fig:2}(b),
at the $\,\order{E^2}\,$ and to all orders in $\,(\xi/\Mp)$\,,
\beqa
\label{eq:T-higgs-s}
\T_\phi^s(\pi^+\pi^- \!\to \pi^0\pi^0) ~=\,
-\FR{\,36\zeta^2\xi^4v^2\,}{\Mp^4}E^2+\order{E^0} \,.
\eeqa
Thus, summing up \eqref{eq:WWZZgoldstone} and \eqref{eq:T-higgs-s}, we have the
full Goldstone amplitude at $\,\order{E^2}\,$,
\beqa
\label{eq:pipi+-00}
\T (\pi^+\pi^- \!\to \pi^0\pi^0) ~=~ \T_h^s + \T_\phi^s
~=~ \bigg[\FR{\,6\xi^2\!+ 2\xi\,}{\Mp^2}-\FR{36\xi^4v^2}{\Mp^4}\bigg]E^2
     + \order{E^0} \,,
\eeqa
where we only keep $\xi$-dependent terms which dominate the amplitude for $\,|\xi|\gg 1\,$.\,
The Goldstone amplitude (\ref{eq:pipi+-00}) agrees well with that of the longitudinal
scattering in Eq.\,(\ref{TWWZZ_Jor_Lon}).
Hence, this explicitly demonstrates the longitudinal-Goldstone boson equivalence theorem
with nonminimal coupling $\,\xi\,$ in the Jordan frame.

\begin{figure}[h]
\begin{center}
\vspace*{2mm}
\includegraphics[width=0.95\textwidth]{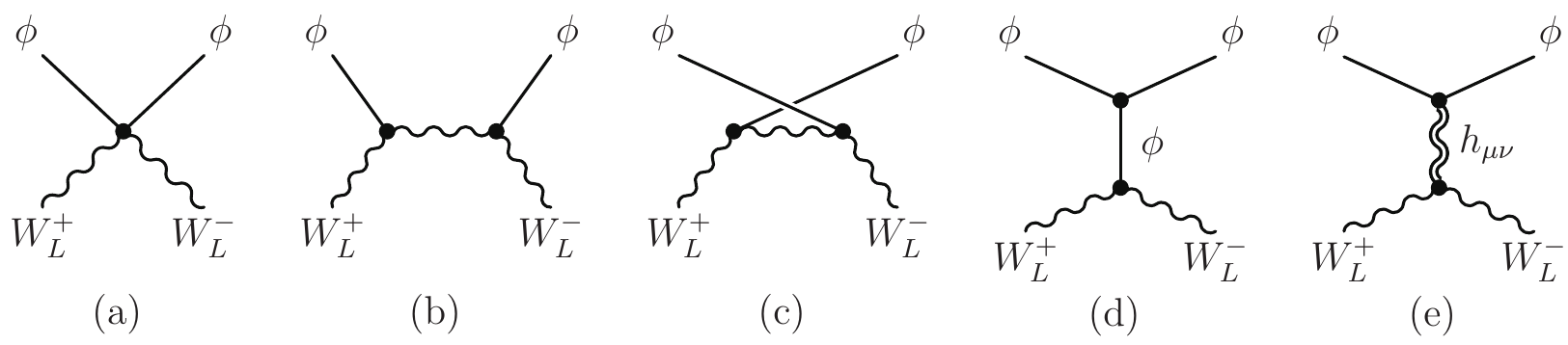}
\vspace*{-1.5mm}
\caption{Longitudinal weak boson scattering process
  $\,W_L^+W_L^- \!\to \phi\phi\,$ in the presence of nonminimal coupling $\,\xi\,$,\,
where the last diagram arises from the graviton exchange in Jordan frame.}
\label{fig:3}
\end{center}
\end{figure}

The second type of processes involves at least a pair of Higgs bosons in the external states.
For illustration, we consider the scattering $\,W_L^+W_L^- \!\to \phi\phi$\,.\,
In unitary gauge, the tree-level contributions to this process are shown in
Fig.\,\ref{fig:3}. We see that graviton also plays a role,
because its coupling with the Higgs boson $\phi$ receives large $\xi$-enhancement
directly from the nonminimal coupling term.

Using the Feynman rules of Appendix\,\ref{A:FRJ},
we evaluate the relevant diagrams in Fig.\,\ref{fig:3},
\beqs
\begin{align}
  \text{Fig.\,3(a)}
  ~=\, &~\Big(1-\FR{4\xi v^2}{\Mp^2}\Big)\zeta^2\FR{E^2}{v^2}+\order{E^0}\,,
  \\[1.5mm]
  \text{Fig.\,3(b)+(c)} ~=\, &-\Big(1-\FR{2\xi v^2}{\Mp^2}\Big)\zeta^2\FR{E^2}{v^2} + \order{E^0}\,,
  \\[1.5mm]
  \text{Fig.\,3(d)} ~=\, &-\FR{\,6\xi^2\!-\xi\,}{\Mp^2}\zeta^4E^2
  +\order{E^0} \,,
  \\[1.5mm]
  \text{Fig.\,3(e)} ~=\, &-\FR{\,\xi+\order{\xi^0}\,}{\Mp^2}\zeta^2E^2
  +\order{E^0} \,.
\end{align}
\eeqs
Summing up all these diagrams, we deduce the scattering amplitude at $\,\order{E^2}$,\,
including all leading terms of $\,(\xi/\Mp)^{2n}$\,,
\bge
\label{TWWss_J}
\begin{aligned}
  \T(W_L^+W_L^- \!\to \phi\phi)
  ~=& -\zeta^2\bigg[1+\FR{3\xi v^2}{\Mp^2}-\Big(1+\FR{\xi v^2}{\Mp^2}\Big)\zeta^2\bigg]\FR{E^2}{v^2}+\order{E^0}
  \\[1mm]
  ~=& -\bigg[\FR{6\xi^2+2\xi}{\Mp^2}-\FR{72\xi^4v^2}{\Mp^4}\bigg]E^2+\order{E^0,\Mp^{-6}} \,,
\end{aligned}
\ede
where we have also retained the first subleading term of $\,\order{\xi/\Mp^{2}}\,$
as before.

It is interesting to note that the above amplitude \eqref{TWWss_J} coincides with
that of $\,W_L^+W_L^-\to Z^0_LZ^0_L$\, in \eqref{TWWZZ_Jor_Lon} at the leading order
of $\,\Mp^{-2}\,$, except to an overall minus sign.
This becomes more transparent when we check the corresponding Goldstone scattering amplitude
of $\,\pi^+\pi^-\to \phi\phi$\,.\,
Similar to $\,\pi^+\pi^-\to \pi^0\pi^0$,\, only the diagram with $s$-channel graviton exchange
contributes at $\,\order{E^2/\Mp^2}$,\, with the final state $\,\pi^0\pi^0$\,
replaced by $\,\phi\phi\,$ in Fig.\,\ref{fig:2}.
Applying the Feynman rules of Appendix\,\ref{A:FRJ}, we compute the leading amplitude
at $\,\order{\Mp^{-2}}\,$,
\bge
\label{Tppss_J}
\T(\pi^+\pi^- \!\to \phi\phi)
~=~ \FR{\,6\xi^2\!+2\xi+\order{\xi^0}\,}{\Mp^2}E^2 + \order{E^0, \Mp^{-4}}\,,
\ede
which coincides with (\ref{eq:WWZZgoldstone}) due to the universal coupling of graviton
at leading order of $\,M_{\textrm{Pl}}^{-2}$\,.\,
According to the equivalence theorem \cite{He:1997zm},
\bge
\label{eq:ET}
  \T(W_L^{a_1},\cdots ,W_L^{a_n},X) ~=~ \T(-\ii\pi^{a_1},\cdots,-\ii\pi^{a_n},X)
   + O(m_W^{}/E_j^{}\,\text{suppressed}) \,,
\ede
where $X$ represents other physical on-shell states.
Hence, for the $2\to 2$ scattering,
a difference of overall minus sign between (\ref{TWWss_J}) and (\ref{Tppss_J}) is expected,
due to the factor $\,(-i)^2=-1\,$ associated with the two external Goldstone fields
$\,\pi^+\pi^-\,$ on the right-hand-side of \eqref{eq:ET}.

For all other processes, we find full agreement between the longitudinal scattering amplitudes
and the corresponding Goldstone scattering amplitudes by explicit calculations,
for $\,E^2\gg m_W^2\,$.\, This justifies the longitudinal-Goldstone boson equivalence theorem
in the presence of Higgs-gravity interactions with nonzero $\,\xi\,$ coupling.
In the following, we summarize all the $\,\order{E^2}\,$ Goldstone amplitudes
for the electrically neutral channels, which include all leading terms of $\,(\xi/\Mp)^{2n}$.\,
We will also retain the first subleading term of $\,\order{\xi/\Mp^{2}}\,$.\,
Thus, we arrive at,
\beqs
\label{GoldAmp}
\begin{eqnarray}
  \T(\pi^+\pi^- \!\to \pi^+\pi^-) &=&
  \bigg[1-\Big(1-\FR{2\xi v^2}{\Mp^2}\Big)\zeta^2\bigg]\FR{(1\!+\cos\theta)}{2v^2}E^2
\n\\
  & \simeq & \bigg[\FR{\,3\xi^2\!+\xi\,}{\Mp^2}-\FR{18 \xi^4v^2}{\Mp^4}\bigg](1\!+\cos\theta)E^2,
\\[2mm]
  \T(\pi^+\pi^- \!\to \pi^0\pi^0) &=& \bigg[1-\Big(1-\FR{2\xi v^2}{\Mp^2}\Big)\zeta^2\bigg]\FR{E^2}{v^2}
\n\\
  & \simeq & \bigg[\FR{\,6\xi^2\! + 2\xi\,}{\Mp^2}-\FR{36\xi^4v^2}{\Mp^4}\bigg]E^2,
\\[2mm]
  \T(\pi^0\pi^0 \!\to \pi^0\pi^0) &=& \order{E^0}\,,
\\[2mm]
  \T(\pi^+\pi^- \!\to \phi\phi) &=&
 \bigg[1+\FR{3\xi v^2}{\Mp^2}-\Big(1+\FR{\xi v^2}{\Mp^2}\Big)\zeta^2\bigg]\zeta^2\FR{E^2}{v^2}
\n\\
  & \simeq & \bigg[\FR{6\xi^2\!+2\xi}{\Mp^2}-\FR{72\xi^4v^2}{\Mp^4}\bigg]E^2,
\\[1.5mm]
  \T(\pi^0\pi^0 \!\to \phi\phi) &=&
  \bigg[1+\FR{3\xi v^2}{\Mp^2}-\Big(1+\FR{\xi v^2}{\Mp^2}\Big)\zeta^2\bigg]\zeta^2\FR{E^2}{v^2}
\n\\
  & \simeq & \bigg[\FR{6\xi^2\!+2\xi}{\Mp^2}-\FR{72\xi^4v^2}{\Mp^4}\bigg]E^2,
\\[2mm]
  \T(\pi^0 \phi \to \pi^0 \phi) &=&
-\bigg[1+\FR{3\xi v^2}{\Mp^2}-\Big(1+\FR{\xi v^2}{\Mp^2}\Big)\zeta^2\bigg]
\FR{(1\!-\cos\theta)}{2v^2}\zeta^2E^2
\n\\
& \simeq & -\bigg[\FR{3\xi^2\!+\xi}{\Mp^2}-\FR{36\xi^4v^2}{\Mp^4}\bigg](1\!-\cos\theta)E^2 ,
\\[2mm]
  \T(\phi\phi \to \phi\phi) &=& \order{E^0} \,.
\end{eqnarray}
\eeqs
These serve as highly nontrivial self-consistency checks of the scattering amplitudes in Jordan frame.
The above can be compared with our previous results
in Einstein frame \cite{XRH}, which were computed at $\,\order{\Mp^{-2}}\,$ only.
Also, we note that
Ref.\,\cite{atkins2} studied linearized gravity in the presence of a nonminimal coupling term
$\,\xi R\phi^2$ (with singlet scalar $\phi$\,) in Jordan frame.
They calculated scattering amplitudes for external spin-(0,\,1/2,\,1)
particles from graviton-exchange at $\,\order{\Mp^{-2}}\,$,\,
where the singlet scalar $\,\phi\,$ has no VEV.
From our results in (\ref{GoldAmp}) with the SM Higgs doublet,
we find that the leading $\xi$ contributions at $\,\order{\Mp^{-2}}\,$
do not depend on the Higgs VEV.
Thus, for the singlet scalar scattering process like $\,ss\to s's'$,\,
our result reduces to that of Ref.\,\cite{atkins2} at $\,\order{\Mp^{-2}}\,$.

\vspace*{2mm}
\subsection{Analysis in Einstein Frame}
\vspace*{2mm}

In this subsection, we proceed with the analysis in Einstein frame.
To demonstrate the nontrivial difference of the current analysis from that in Jordan frame,
we will present explicit calculations for the two processes considered in the previous subsection.

We first compute the amplitude of $\,W_L^+W_L^-\to Z_L^0Z_L^0$\, with the Lagrangian (\ref{DL-int_pg}).
In unitary gauge, the contribution to the scattering amplitudes with positive power of scattering energy $E$
comes from the same diagrams as in Jordan frame in Fig.\;\ref{fig:1}.
As shown in Appendix\,\ref{A:FR},
the $\xi$ corrections to the cubic Higgs-gauge-boson couplings are the same
in both frames. Hence, the amplitude from Fig.\,\ref{fig:1}(d) remains intact, and the sum of the
four diagrams should equal that of (\ref{TWWZZ_Jor_Lon}) in the Jordan frame.
In parallel, we consider the corresponding Goldstone boson scattering process
$\,\pi^+\pi^-\to \pi^0\pi^0$\,.\,
Here, the deviation from the SM is from higher dimensional operators in the
Lagrangian (\ref{DL_int_ss}). Using the Feynman rules in Appendix\,\ref{EFfeynmanrule},
we find that the leading amplitude for this process arises from the contact interaction,
\bge
\label{eq:EF+-00}
  \mathcal{T}(\pi^+\pi^- \!\to \pi^0\pi^0)
  ~=~ \[\FR{\,6\xi^2\!+2\xi\,}{\Mp^2} -\FR{36\xi^4v^2}{\Mp^4}\]\! E^2
       + \order{E^0, \Mp^{-6}} \,,
\ede
which coincides with the amplitude of $\,W_L^+W_L^-\to Z_L^0Z_L^0$\,.\,
This shows that the longitudinal-Goldstone boson equivalence theorem holds for this process
in Einstein frame, and the derived scattering amplitudes in both frames are consistent.
For the sake of later physical analysis, we present the amplitudes of
$\,W_L^\pm W_L^\pm \to W_L^\pm W_L^\pm$\, and $\pi^\pm\pi^\pm\to\pi^\pm\pi^\pm$\, as well.
By crossing symmetry, we infer from \eqref{eq:EF+-00},
\beqa
\label{eq:EF++++}
  \mathcal{T}(W_L^\pm W_L^\pm \!\to W_L^\pm W_L^\pm )
  & \simeq &  \mathcal{T}(\pi^\pm\pi^\pm \!\to\pi^\pm\pi^\pm )
\n\\[1mm]
  & = & -\[\FR{\,6\xi^2\!+2\xi\,}{\Mp^2} -\FR{36\xi^4v^2}{\Mp^4}\]\!\! E^2
       + \order{E^0, \Mp^{-6}} \,.
\eeqa

Next, we analyze the other scattering process $\,W_L^+W_L^-\to \phi\phi$\,.\,
Different from the case in Jordan frame, no $\xi$-contribution arises from the graviton-exchange
diagram in Fig.\,\ref{fig:3}(e).
Hence, only the first four diagrams of Fig.\,\ref{fig:3} are relevant to our study in Einstein frame.
With the Feynman rules of Appendix\,\ref{EFfeynmanrule},
we find a difference for the diagram of Fig.\,\ref{fig:3}(a)
between the two frames, while the diagrams (b), (c) and (d) remain intact.
Despite such a difference, we have verified the following equality at $\,\order{E^2}\,$,
\begin{eqnarray}
\label{WWphiphiJFEF}
\left.\[\text{Fig.\,\ref{fig:3}(a)}+\text{Fig.\,\ref{fig:3}(e)}\]\right|_{\textrm{Jordan}}
~=~
\left.\text{Fig.\,\ref{fig:3}(a)}\right|_{\textrm{Einstein}}  \,.
\end{eqnarray}
Hence, the amplitude fully coincides with that in the Jordan frame.
This explicitly demonstrates the equivalence between the two frames via the above scattering process.
For the corresponding Goldstone boson scattering, it receives contributions from contact interaction
as well as $s$-channel Higgs exchange. We find the same result as (\ref{Tppss_J}) at the leading order.

Then, we systematically extend the above calculations to all other scattering channels.
We reveal that for both the longitudinal gauge boson scattering and Goldstone boson scattering,
the results coincide with those in the Jordan frame.
As an advantage of the Einstein frame analysis, it is easier to extract the $\xi$-dependent $E^2$-terms
without invoking tedious calculations with the graviton-exchange.
Here we present the amplitudes at $\order{E^2}$,\, keeping all the leading power terms of
$\,(\xi/\Mp)^{n}$\,,\, but dropping the subleading terms in which the power of $\,\xi\,$ is lower than
that of $(\xi/\Mp)^{n}$\, at each given order of $\,1/\Mp^{n}\,$.\,  We find that so long as $\,\xi>\order{1}$\,,\,
this always gives a good approximation for our unitarity analysis in the next subsection.
\begin{eqnarray}
  \label{GoldAmpGeneral}
  \T(\pi^+\pi^- \!\to \pi^+\pi^-) &=&~ \FR{\,(1-\zeta^2)\,}{2v^2}(1\!+\cos\theta)E^2 \,,
\n\\[2mm]
  \T(\pi^+\pi^- \!\to \pi^0\pi^0) &=&~ \FR{\,(1-\zeta^2)\,}{v^2}E^2 \,,
\n\\[2mm]
\T(\pi^0\pi^0 \!\to \pi^0\pi^0) &=&~ \order{E^0} \,,
\\[2mm]
  \T(\pi^+\pi^- \!\to \phi\phi) &=&~ \FR{\,(1-\zeta^2)\zeta^2\,}{v^2}E^2 \,,
\n\\[2mm]
  \T(\pi^0\pi^0 \!\to \phi\phi) &=&~ \FR{\,(1-\zeta^2)\zeta^2\,}{v^2}E^2 \,,
\n\\[2mm]
\T(\pi^0 \phi \to \pi^0 \phi) &=& -\FR{\,(1-\zeta^2)\zeta^2\,}{2v^2}(1\!-\cos\theta)E^2 \,,
\n\\[2mm]
  \T(\phi\phi \to\phi\phi) &=&~ \order{E^0} \,.
\n
\end{eqnarray}
We note that the \,$(\xi/\Mp)^n$\, terms originate from the $\xi$-enhanced effective operators
and the rescaling factor $\,\zeta$\,.\,
As a consistency check, we note that the above expressions reduce to (\ref{GoldAmp}) at $\order{\xi^2/\Mp^2}$ and $\order{\xi^4/\Mp^4}$ under the $\,1/\Mp\,$ expansion.

So far, as a by product,
we have explicitly demonstrated the equivalence between Jordan frame and Einstein frame
for high energy scattering of weak bosons and Higgs bosons under the flat background metric.
There are debates on the physical (in)equivalence between Jordan and Einstein frames
in the literature \cite{conformal,Bezrukov:2009db,Barvinsky:2009fy,Barvinsky:2009ii,JE-new}.
Our study supports the equivalence of the two frames
from the tree-level analysis of weak boson and Higgs boson scatterings
(which were not considered before).
The reason is that we can take the spacetime be flat in both frames,
and then the frame transformation is just a nonlinear field redefinition. According to the
theorem \'{a} la CCWZ \cite{CCW}, a field redefinition $\,\phi\to f(\phi)\,$
will not change the on-shell $S$-matrix involving $\,\phi$\,,\,
provided that the nonlinear local transformation $\,f(\phi)\,$ has the form
$\,f(\phi)=\phi F(\phi)\,$ where $\,F(\phi)\,$ is another local function of $\,\phi\,$ satisfying
$\,F(0)=1\,$.\,  This is exactly the case for the canonical degrees of freedom in the two frames,
as we explicitly demonstrated in (\ref{eq:EFfieldRedef}). Hence, we should expect
the same results for high energy weak boson scatterings in both frames.

In passing, we note that the equivalence of the two frames is more involved beyond
tree-level, as pointed out before
\cite{Barvinsky:2008ia,Bezrukov:2009db,Barvinsky:2009fy,Barvinsky:2009ii,JE-new}.
Due to the quantum anomaly from the Weyl transformation between the two frames
and the unknown UV dynamics of quantum gravity,
the loop analysis and renormalization prescriptions would suffer ambiguity.
For example, there are different ways of choosing renormalization scale $\mu$,\,
concerning whether $\mu$ is field-independent
in Einstein frame or in Jordan frame \cite{Bezrukov:2009db}.
A possible prescription of realizing the frame-equivalence at quantum level was
recently discussed in \cite{JE-new}.

 Before concluding this section, we make further remarks on the physical impact of the leading order
 $\,\xi\,$ corrections to the scattering amplitudes (\ref{GoldAmp}).
 At the first nontrivial order of $\,\Mp^{-2}$,\, we have both $\,\xi^2\,$ and $\,\xi\,$
 contributions to the amplitudes, where the $\,\xi^2\,$ terms are dominant for $\,|\xi|\gg 1\,$.\,
 Its impact can be classified into three categories.
 The first one is a universal suppression factor $\,\zeta < 1\,$ for any coupling involving
 the Higgs field, such as the Higgs boson production process\,\cite{atkins}.
 The second class of $\xi^2$-dependent processes are the weak boson scatterings
 which we study in this paper. We have analyzed the {\it anomalous} cubic Higgs-gauge coupling
 and quartic scalar couplings in Einstein frame, and computed the graviton-exchanges in Jordan frame.
 We find that they cause non-canceled $\,\xi^2\,$ (and $\xi$) dependent $E^2$-contributions
 in the longitudinal and Goldstone boson scattering amplitudes,
 which can become significant as the increase of scattering energy $\,E$\,.\,
 Hence, the longitudinal $WW$ scattering can provide sensitive probe of $\,\xi\,$ coupling
 via energy-enhanced leading contributions of $\,\order{\xi^2E^2/\Mp^2}$.\,
 The third class of $\xi^2$-involved processes are those containing the cubic Higgs self-coupling.
 As shown in (\ref{DL_int_ss}), such processes will also be enhanced at high energies
 by the $\xi^2$-dependent derivative cubic Higgs couplings.
 The future high energy $pp$ colliders ($50-100$\,TeV) \cite{FCC}
 will further probe such anomalous cubic Higgs couplings. Finally, the $\,\order{\xi/\Mp^2}$\, terms
 arise from the Weyl factor $\,\Omega$\,,\, including bosonic and fermionic couplings of the Higgs boson.
 But they are negligible relative to the leading contributions of $\,\order{\xi^2/\Mp^2}$\, for
 $\,|\xi|\gg 1$\,.

\vspace*{2mm}
\subsection{Perturbative Unitarity Bound on Higgs-Gravity Coupling}
\label{sec3.3}
\vspace*{2mm}

In the above two subsections, we have derived the high energy weak boson scattering amplitudes
in the Jordan and Einstein frames,
and demonstrated the longitudinal-Goldstone boson equivalence theorem in both frames.
We further showed that the two frames give the same scattering amplitudes.
With these, we will derive perturbative unitarity bound on
the Higgs-gravity coupling $\,\xi\,$ via coupled channel analysis in this subsection.

Given the scattering amplitudes in the previous subsections,
we compute the partial wave amplitudes for the Goldstone and Higgs bosons,
\beqa
a_\ell^{}(E) ~=~
\FR{1}{32\pi}\int_{-1}^1\!\!\di\cos\theta\,
\mathrm{P}_\ell^{}(\cos\theta)\,\mathcal{T}(E,\theta) \,.
\eeqa
For the present case, the partial wave amplitudes form a $4\times 4$ matrix
among the four initial/final states,
$|\pi^+\pi^-\ra$, $\fr{1}{\sqrt 2}|\pi^0\pi^0\ra$,
$\fr{1}{\sqrt 2}|h^0h^0\ra$, and $|\pi^0 h^0\ra$.
In coupled channel analysis, we will impose the $s$-wave unitarity condition,
$\,|\mathfrak{Re}a_0^{}| < 1/2$\,,\,  on the maximal eigenvalue of the matrix
$\,a_0^{}$\,.

 \begin{figure}[t]
 \begin{center}
 \includegraphics[width=0.495\textwidth]{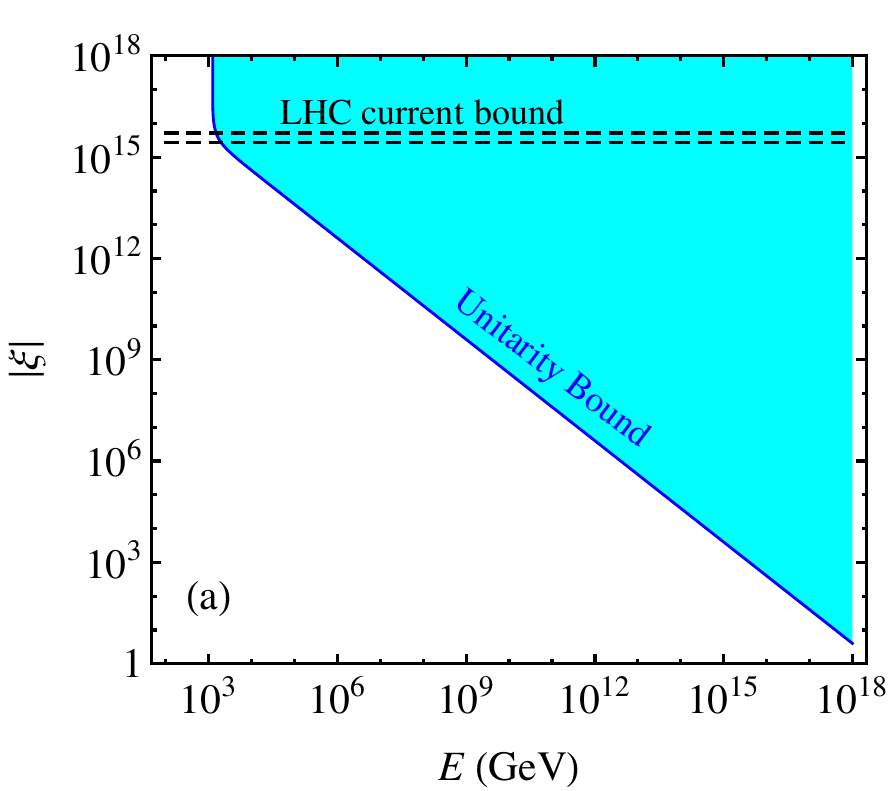}
 \includegraphics[width=0.495\textwidth]{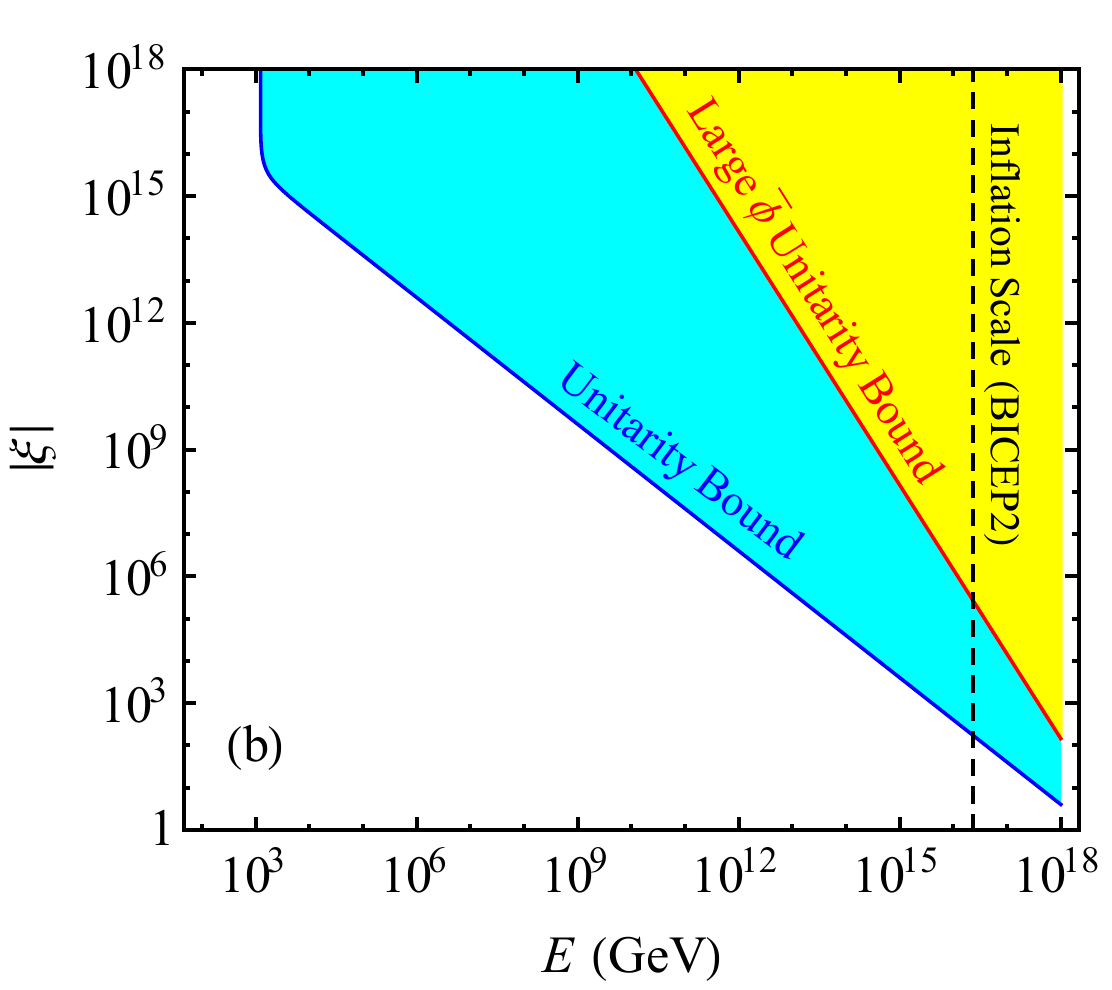}
 \vspace*{-1mm}
 \caption{Perturbative unitarity bound on the Higgs-gravity coupling $\,\xi\,$ as a function of
 center-of-mass energy $\,E\,$.\, In plot-(a), the blue curve denotes the bound derived at $\,\phib =v\,$,\,
 where the shaded region violates perturbative unitarity.
 In plot-(b), we further present the unitarity bound at large backround field
 $\,\bar\phi=\Mp/\sqrt{\xi}\,$ (red curve),
 as compared to the unitarity bound at $\,\phib =v\,$ (blue curve).
 The shaded yellow area above the red curve violates perturbative unitarity,
 and its physical implications will be explained in Sec.\,\ref{sec:4.3}.
 The vertical dashed line denotes the inflation scale ($2.3\times 10^{16}$\,GeV)
 as indicated by the BICEP2 data\,\cite{BICHEP2}.
 For comparison, the upper (lower) horizontal dashed line in plot-(a) denotes the
 $3\sigma$ bound $|\xi|<5.2\times 10^{15}$ (\,$|\xi|<2.7\times 10^{15}$\,) derived from
 the current CMS (ATLAS) Higgs data\,\cite{LHC2013} at the LHC.}
 \label{fig:4}
 \end{center}
 \vspace*{-4mm}
 \end{figure}

From (\ref{GoldAmpGeneral}), we deduce the following $s$-wave amplitude at $\,\order{E^2}\,$
and to all orders in $\,(\xi/\Mp)\,$,
\begin{align}
  a_0^{}(E) ~=\, \FR{\,(1-\zeta^2)E^2\,}{32\pi v^2}\!
  \begin{pmatrix}
    1 & \sqrt2 & \sqrt2\zeta^2 & 0 \\[1mm]
    \sqrt2 & 0 & \zeta^2 & 0 \\[1mm]
    \sqrt2 \zeta^2 & \zeta^2 & 0 & 0 \\[1mm]
    0 & 0 & 0 & -\zeta^2
  \end{pmatrix} \!.~~~~~
\label{a0-4x4}
\end{align}
Thus, we further derive its eigenvalues,
\begin{align}
a_0^{\text{diag}}(E) ~=~
\FR{(1\!-\zeta^2)E^2}{32\pi v^2}
\text{diag}\Big(1\!+\!\sqrt{1\!+3\zeta^4},\, 1\!-\!\sqrt{1\!+3\zeta^4},\, -\zeta^2,\, -1\Big) \,.
\label{a0-diag}
\end{align}

Imposing the partial wave unitarity condition on the maximal eigenvalue of (\ref{a0-diag}),
we deduce the following perturbative unitarity bound on the scattering energy $\,E$\,
for a given value of the $\,\xi\,$ coupling,
\beqa
\label{eq:UB-E}
E  ~<\,
\frac{\sqrt{16\pi\,} v}
{\,\big[(1-\zeta^2)\big(1 \!+\!\sqrt{1\!+3\zeta^4}\big)\big]^{1/2}\,} \,.
\eeqa
We can translate this into a constraint on the Higgs-curvature coupling $\,\xi\,$
at a given energy scale $\,E\,$,
\beqa
\label{eq:UB-xi0}
(1\!-\zeta^2)\big(1\!+\!\sqrt{1\!+\!3\zeta^4}\,\big) ~<~ \frac{\,16\pi v^2\,}{E^2} \,.
\eeqa
For most of the parameter range of physical interest, we find that the expansion of $\,\zeta\,$
in terms of $\,\xi v/\Mp^{}\,$ gives a good approximation.
Thus, from (\ref{eq:UB-xi0}) or \eqref{eq:UB-E},
we may derive a simplified perturbative unitarity bound on
the $\,\xi\,$ coupling,
\beqa
\label{eq:UB-xi}
  |\xi| ~<~  \FR{\sqrt{8\pi}\Mp^{}\,}{3E}\bigg[1+\FR{4\pi v^2}{E^2}+\mathcal{O}\Big(\FR{v}{E}\Big)^4\bigg] \,.
\eeqa
We see that the expansion in the brackets of the right-hand-side of \eqref{eq:UB-xi}
is practically in terms of the ratio $\,v^2/E^2\,$.\, Hence, the approximate formula
\eqref{eq:UB-xi} works well so long as $\,E> \order{1\text{TeV}}\,$.\,
We also note that the tree-level unitarity violation is a signal of requiring higher order
nonperturbative effects and the inclusion of new resonance(s) in the effective field theory.
As we clarified before \cite{XRH}, the perturbative unitarity bounds\,\cite{UB0,UB1,LQT}
are fully justified and important. Especially, the unitarity bound on the SM Higgs boson mass
$\,m_h^{} < \sqrt{8\pi/3}\,v\simeq 712\,$GeV \cite{LQT}
has been well supported by the recent LHC Higgs discovery\,\cite{LHC2012,LHC2013}
with $\,m_h^{}\simeq 125\,\text{GeV}\, (<712\,\text{GeV})\,$.

In Fig.\,\ref{fig:4}(a), we present the perturbative unitarity bound of $\,\xi\,$ as a function
of scattering energy $E$, up to $\,E = 10^{18}\,\text{GeV}\lesssim\Mp$, by the thick blue curve,
where we set $\,\bar\phi=v\,$ for the electroweak vacuum.
It is useful to check a special limit of $\,\xi\to\infty\,$ for the
condition \eqref{eq:UB-xi0}. In this case, we have $\,\zeta\to 0\,$ and the Higgs field decouples.
Thus, the bound \eqref{eq:UB-xi0} reduces to
$\,E < \sqrt{8\pi}v\simeq 1.23\,$TeV.\, This bound is nicely reflected by the
asymptotical behavior of the blue curve around $\,E=\order{1\text{TeV}}\,$,\,
as depicted in Fig.\,\ref{fig:4}(a).
For the application to Higgs inflations in the next section, we also show the unitarity limit
(thick red curve) for the inflation background $\,\bar\phi=\Mp/\sqrt{\xi\,}\,$
in Fig.\,\ref{fig:4}(b), which is significantly relieved than the small field bound
at $\,\bar\phi=v\,$.\, We will discuss this further in Sec.\,\ref{sec:4.3}.

Fig.\,\ref{fig:4}(a) shows that the unitarity bound puts highly nontrivial constraints on the
Higgs-curvature coupling $\,\xi\,$ in the perturbative formulation.
For the effective theory of the SM\,+\,GR with Planck mass
$\,\Mp\,$ as the UV cutoff, the weak boson scattering energy can reach up to
$\,E = 10^{17-18}\,\text{GeV} < \Mp$.\,
In this case, we find that the perturbative unitarity bound in plot-(a) places
stringent new limits, $\,\xi \lesssim \order{10-1}\,$,\,
for $\,\bar\phi=v\,$.\,
Besides, we would like to note that Atkins and Calmet\,\cite{atkins} derived
an interesting bound on $\,\xi\,$ from the 2012 LHC data \cite{LHC2012}.
The latest update of the LHC analyses \cite{LHC2013} leads to the
Higgs signal strengths, $\,\hat{\mu}=1.30\pm 0.20\,$ [ATLAS] and
$\,\hat{\mu}= 0.80\pm 0.14\,$ [CMS].
From this, we have the refined $3\sigma$ upper limits,  $\,|\xi|<2.7\times 10^{15}$\, [ATLAS]
and $\,|\xi| < 5.2\times 10^{15}$\, [CMS].
For comparison, we plot the ATLAS and CMS limits in Fig.\,\ref{fig:4}(a)
by the horizontal lower and upper dashed lines, respectively.
For the electroweak vacuum $\,\bar\phi=v\,$,\,
we see that once the scattering energy $E$ exceeds
$\,\order{\text{TeV}}\,$,\, the perturbative unitarity bound becomes much more stringent.

\vspace*{2mm}
\subsection{Probing Higgs-Gravity Coupling via Weak Boson Scattering}
\vspace*{2mm}

It is possible that Nature may have chosen a lower UV cutoff for
the effective theory of SM\,+\,GR.  Thus, the perturbative unitarity bound
in Fig.\,\ref{fig:4} will allow much larger $\,\xi\,$ values.
An intriguing situation is that the UV cutoff sets in at a scale close to
$\order{\text{TeV}}$,\, say,
$\,\cut = \order{10\text{TeV}}$.\,
Thus, the coupling $\,\xi\,$ can reach\footnote{From the theory side,
we have no preferred natural values for the dimensionless coupling $\,\xi\,$.\,
Note that the $\,\xi\,$ coupling in the Feynman vertices is
always suppressed by the factor $\,v^2/M_{\text{Pl}}^2$\, or $\,E^2/M_{\text{Pl}}^2$\,
[cf.\ \eqref{DL_int_ss}-\eqref{DL-int_pg}].
Hence, a large $\,\xi\,$ coupling is fine so long as it respects the
perturbation expansion (Fig.\,\ref{fig:4}).}
$\,\xi = \order{10^{15}}\,$.\,
Such a relatively low UV cutoff gives a conceptually simple resolution to the
hierarchy problem and makes the SM Higgs sector natural up to\footnote{For the current
effective theory study, we are not concerned with any detail of the UV dynamics above
$\,\cut =\order{10\text{TeV}}\,$.\,
Many well-motivated TeV scale quantum gravity theories exist on the market.
For instance, an extra dimensional model with compactification scale of
$\,\order{10\text{TeV}}\,$
will reveal its Kaluza-Klein modes at energies above this scale.\,}\,
$\,\cut = \order{10\,\text{TeV}}\,$.\,
This will open up an exciting possibility that
the upcoming runs of LHC\,(14\,TeV)
and the future high energy $pp$ colliders ($50-100$\,TeV)\,\cite{FCC}
can effectively probe such Higgs-gravity interactions with $\,\xi = \order{10^{14-15}}\,$
via weak boson scattering experiments.

\begin{figure}[t]
 \begin{center}
 \includegraphics[width=0.54\textwidth]{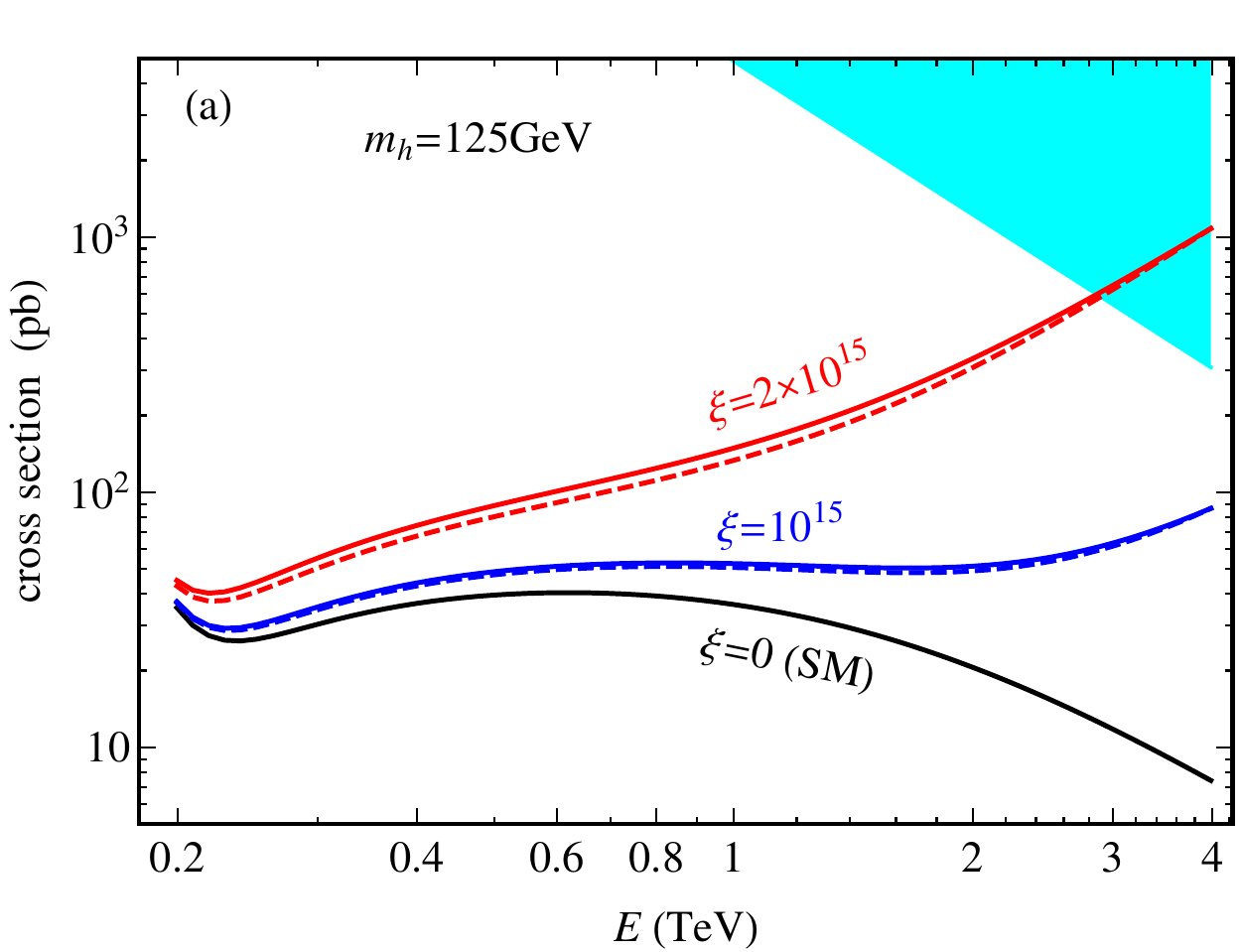}
 \\
 \includegraphics[width=0.49\textwidth]{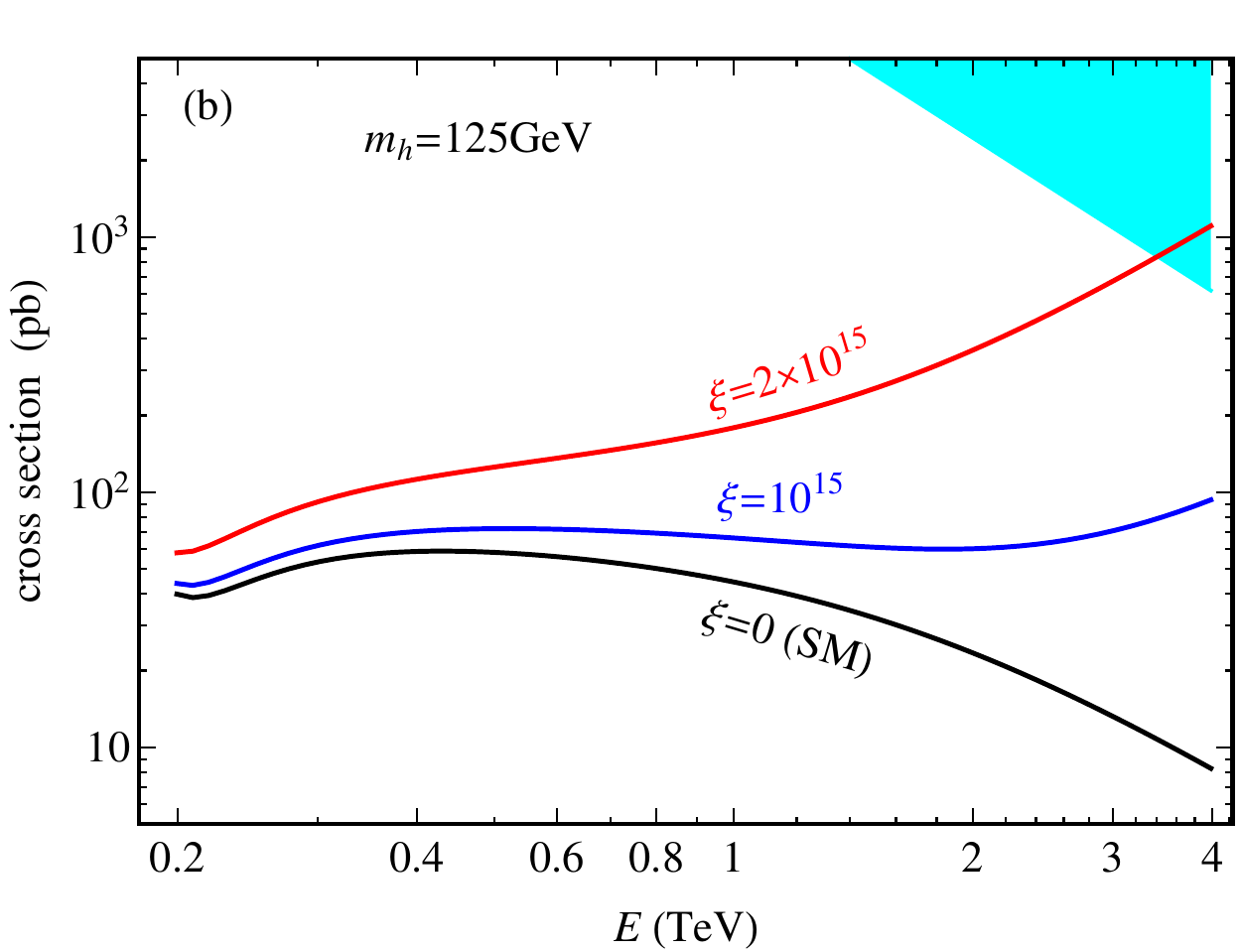}
 \includegraphics[width=0.49\textwidth]{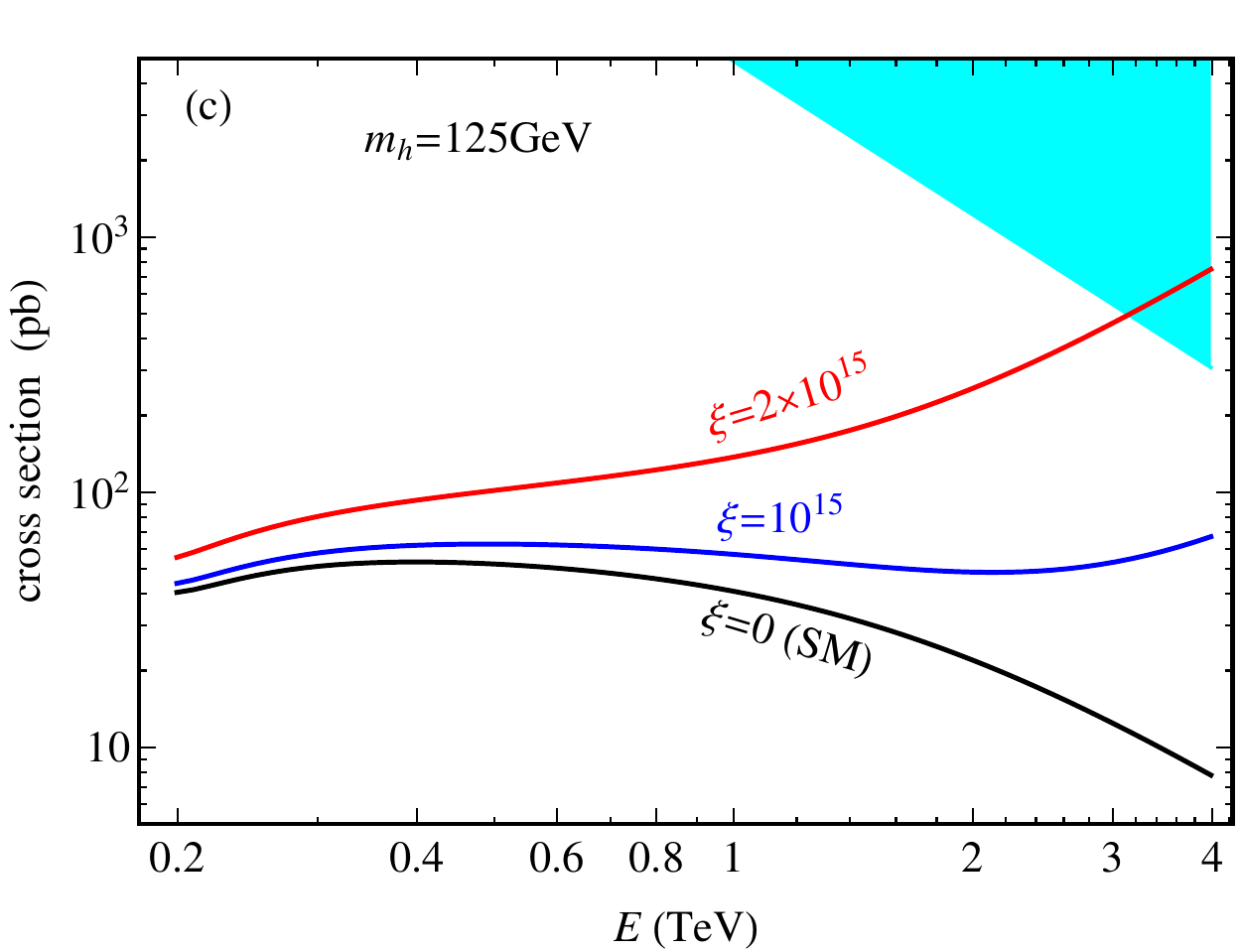}
 \caption{Cross sections of weak boson scattering with Higgs-gravity coupling $\,\xi\,$
 for the relevant energy range at the LHC\,(14\,TeV).
 Plot-(a): $\,W^+_L W^-_L \to Z_L^0Z_L^0$\,.\,
 Plot-(b): $\,W^\pm_L W^\pm_L \to W^\pm_L W^\pm_L\,$.\,
 Plot-(c): $\,W_L^+Z_L^0\to W_L^+Z_L^0$\,.\,
 In each plot, we depict the predictions for
 $\,\xi =(2\times\! 10^{15},\,10^{15},\,0)$\, by the (red,\,blue,\,black) curves.
 The pure SM result ($\xi=0$) is given by the black curve.
 We have input the Higgs boson mass $\,m_h^{}=125\,$GeV.
 The shaded (light-blue) region violates perturbative unitarity.
 The dashed curves in plot-(a) include the RG running
 of $\,\xi\,$,\, with initial values taken at $\,E=4$\,TeV.
 }
 \label{fig:5}
 \end{center}
 \vspace*{-5mm}
 \end{figure}

The weak boson scattering is a crucial experiment for the LHC to
test new physics of electroweak symmetry breaking beyond the SM Higgs sector \cite{WW}.
Hence, we study the weak boson scattering cross sections, and analyze three major processes,
$\,W^+_L W^-_L \to Z_L^0Z_L^0\,$,\, $\,W^\pm_L W^\pm_L \to W^\pm_L W^\pm_L\,$,\,
and $\,W^\pm_L Z^0_L \to W^\pm_L Z^0_L\,$.\,
For our study, we consider the intriguing effective theory of low scale quantum gravity
with two sample UV cutoffs, $\,\cut = \order{10\,\text{TeV}}\,$
or $\,\cut = \order{50\,\text{TeV}}\,$.\,
For tests at the LHC\,(14\,TeV),
we consider the effective theory with $\,\cut = \order{10\,\text{TeV}}\,$.\,
The relevant energy range for the $WW$ scattering at the LHC is around $\,0.2-4\,$TeV.
From Fig.\,\ref{fig:4}(a), we find that for the $WW$ scattering energy $\,E < 4\,$TeV,
the size of the Higgs-curvature coupling can be fairly large, $\,\xi \lesssim O(10^{15})\,$.\,
In Fig.\,\ref{fig:5}(a)-(c), we present the $WW$ scattering cross sections
for two sample inputs,
$\,\xi = 2\times\! 10^{15},\,10^{15}$\,,\, in comparison with
the SM result of $\,\xi =0\,$.\,
In all three plots (a)-(c), we have input the Higgs boson mass
$\,m_h^{}=125\,$GeV based on the LHC data \cite{LHC2012,LHC2013}.
To remove the special kinematical region of $t/u$-channel diagrams around
$\,\theta = 0,\,\pi\,$,\, we add a modest cut $\,|\cos\theta|< 0.995\,$
for all plots.
Furthermore, we also place the unitarity condition on the scattering cross section,
$\,\sigma < 4\pi \rho_e^{}/E^2 \,$ \cite{Dicus:2004rg}.\,
The shaded light-blue region in each plot of Fig.\,\ref{fig:5} violates
perturbative unitarity.
(In this condition, $\,\rho_e^{}\,$ denotes the identical particle factor for
final state of elastic channel, and for inelastic channel $\,\rho_e^{}\,$ is fixed
by the corresponding elastic channel with the same initial
state as the inelastic channel \cite{Dicus:2004rg}.)

With the sample inputs $\,\xi = 10^{15}\,$ and $\,\xi = 2\!\times\! 10^{15}$\, for
Fig.\,\ref{fig:5}(a)-(c), we see that the $WW$ scattering cross sections
exhibit different behaviors and give sizable excesses above the SM expectations (\,$\xi =0$\,).
We note that these non-resonance behaviors are universal and are predicted to show up
in all weak boson scattering channels\,\footnote{For TeV scale quantum gravity via the spontaneous
dimensional reduction approach (with a 125\,GeV non-SM Higgs boson), we found universal behaviors
of non-resonant $WW$ scattering manifested in a different manner \cite{He:2013ub}.
For a light non-SM Higgs boson in the 4d effective Lagrangian (without gravity),
deviations in the non-resonant $WW$ scattering were studied for the LHC before \cite{He:2002qi}.},
in contrast to the traditional new physics models of the electroweak symmetry breaking \cite{WW}.
The upcoming runs of the LHC\,(14\,TeV) and the upgraded high luminosity LHC
will discriminate such distinctive features of the weak boson scattering signals.

 \begin{figure}[t]
 \begin{center}
 \vspace*{-4mm}
 \includegraphics[width=0.54\textwidth]{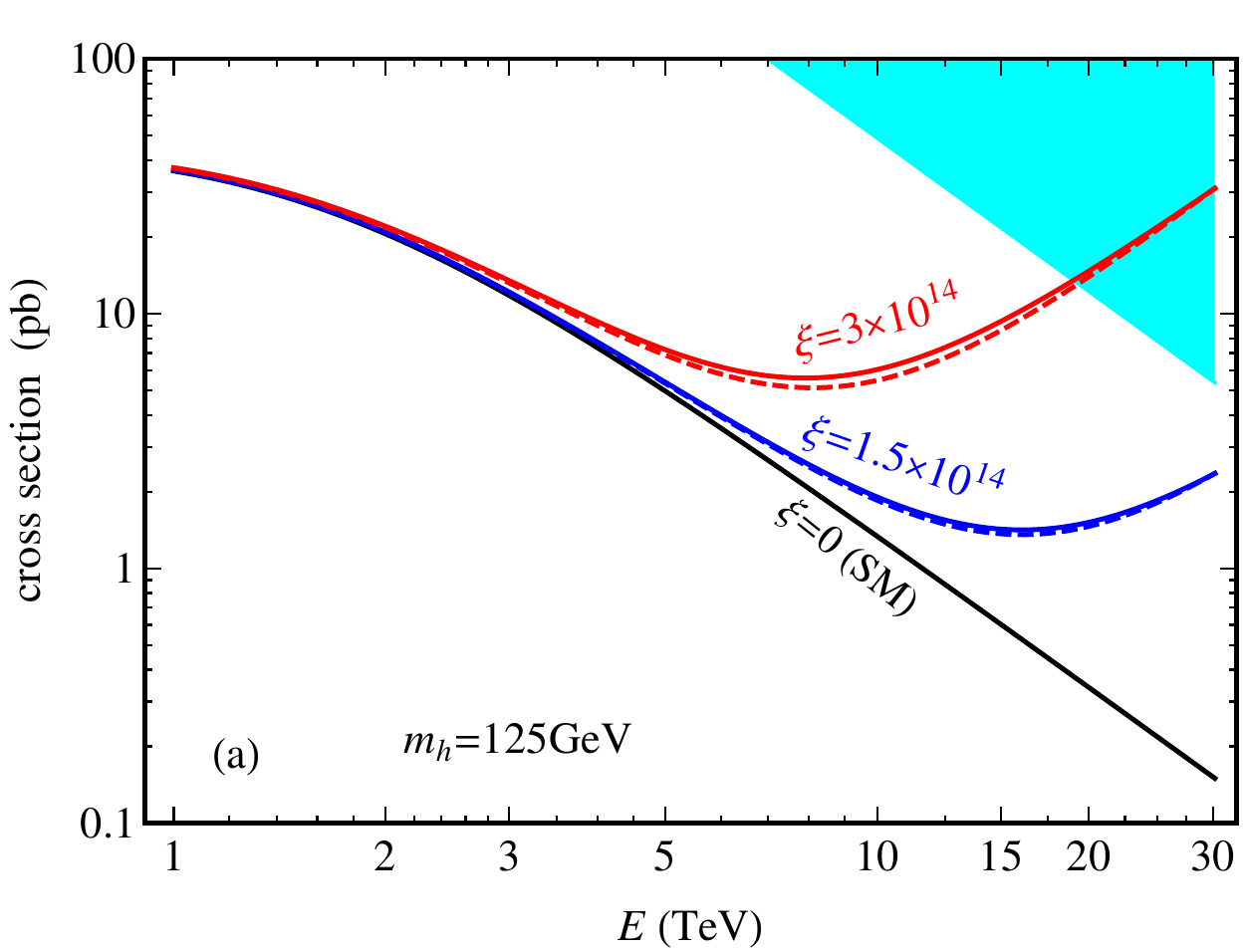}
 \\
 \includegraphics[width=0.49\textwidth]{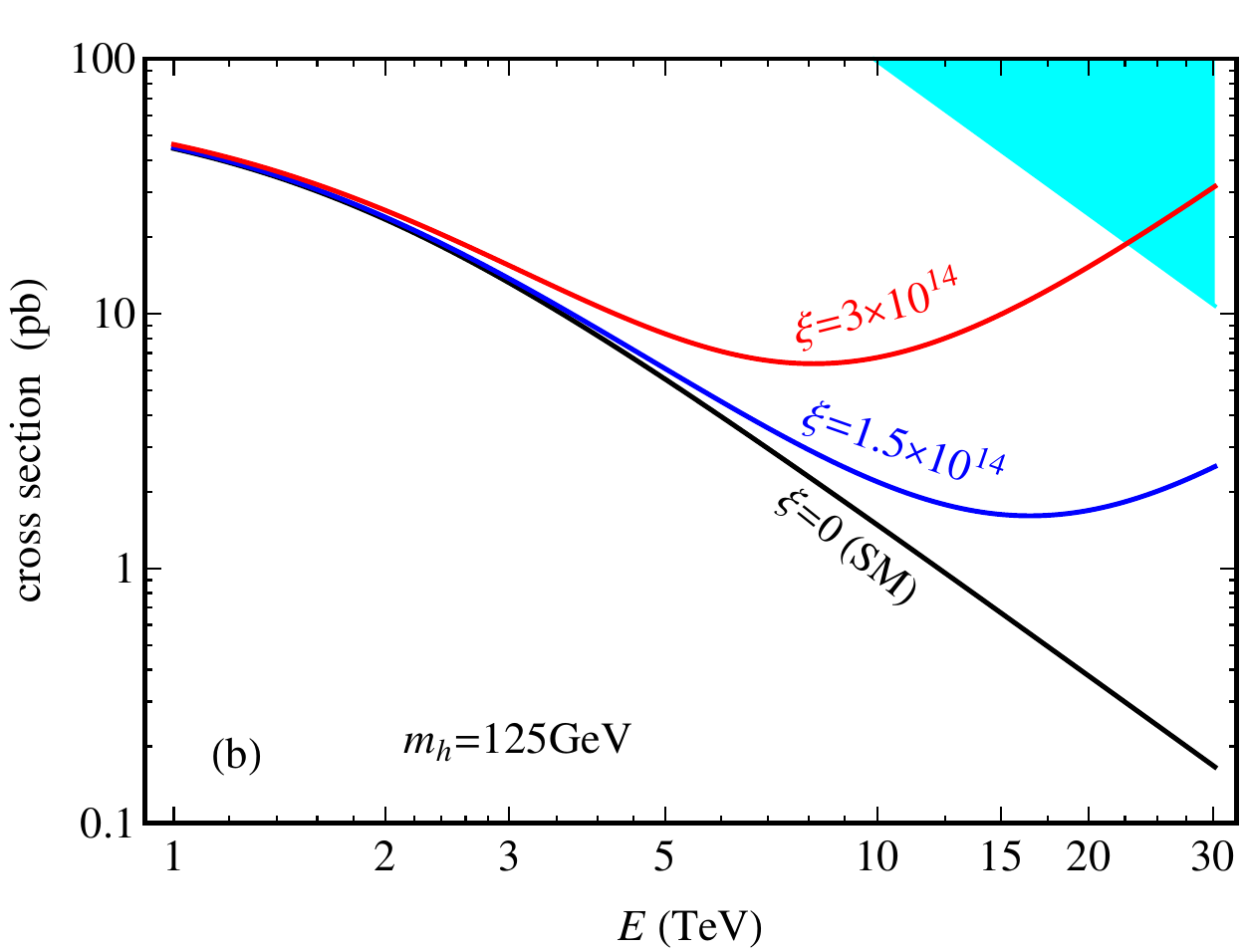}
 \includegraphics[width=0.49\textwidth]{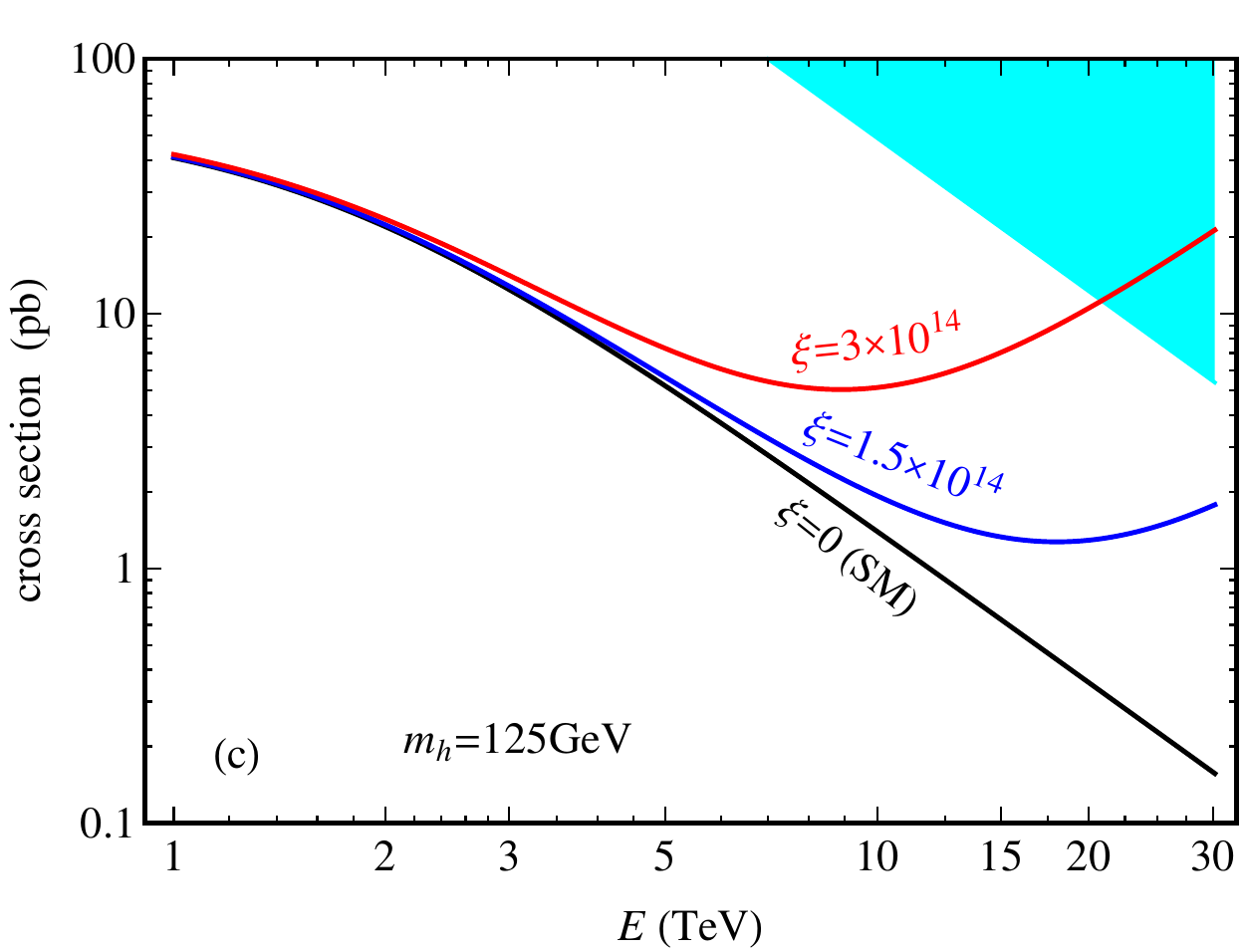}
 \caption{Cross sections of weak boson scattering with Higgs-gravity coupling $\,\xi\,$
 for the relevant energy range at the future $pp$ collider ($50-100$\,TeV).
 Plot-(a): $\,W^+_L W^-_L \to Z_L^0Z_L^0$\,.\,
 Plot-(b): $\,W^\pm_L W^\pm_L \to W^\pm_L W^\pm_L\,$.\,
 Plot-(c): $\,W_L^+Z_L^0\to W_L^+Z_L^0$\,.\,
 In each plot, we depict the predictions for
 $\,\xi =(3\times\! 10^{14},\, 1.5\times\! 10^{14},\,0)$\, by the (red,\,blue,\,black) curves.
 The pure SM result ($\xi=0$) is given by the black curve.
 We have input the Higgs boson mass $\,m_h^{}=125\,$GeV.
 The shaded (light-blue) region violates perturbative unitarity.
 The dashed curves in plot-(a) include the RG running
 of $\,\xi\,$,\, with initial values taken at $\,E=30$\,TeV.
 }
 \label{fig:6}
 \end{center}
 \vspace*{-5mm}
 \end{figure}

For the present analysis, we have also examined renormalization group (RG) running effects of
the $\xi$ coupling. The one-loop RG beta function for $\,\xi\,$ is given by \cite{RGxi},
%
\beqa
\label{eq:RG-xi}
\be (\xi) ~=~ \FR{1+6\xi}{(4\pi)^2}\(2\lam+y_t^2-\FR{3}{4}g^2-\FR{1}{4}g'^2\) \!,
\eeqa
where $(g,\,g')$ are the electroweak gauge couplings, and $(\lambda,\,y_t^{})$
denote the Higg self-coupling and top-quark Yukawa coupling, respectively.
The RG equation \eqref{eq:RG-xi} was derived by treating gravity
as external field \cite{RGxi}. Using quantized metric and including gravitons in the
loop will contribute additional subleading terms of $\,{\cal O}(\xi^0)$,\, as shown in
Ref.\,\cite{Barvinsky:2009fy}. But, for the present study with large values of $\,\xi\gg 1\,$,\,
it is safe to use the approximate RG equation \eqref{eq:RG-xi}, where the leading $\xi$-terms
fully dominate the coupling running.
For numerical analysis of Eq.\,\eqref{eq:RG-xi},
we take initial values $\,\xi= 10^{15}\,$ and $\,2\times 10^{15}$,\,
respectively, at $E=4$\,TeV, and run the
$\,\xi\,$ coupling downward with the $\beta$ function \eqref{eq:RG-xi}. We compute the RG improved
cross sections by including this running $\,\xi\,$ coupling.
We plot the RG improved cross sections in Fig.\,\ref{fig:5}(a) as dashed curves.
We compare them with the pure tree-level cross sections (in solid curves), and find that the corrections
from the RG running of $\,\xi\,$ coupling are fairly small for the relevant LHC energy range.

Next, we study the probe of Higgs-gravity coupling $\,\xi\,$ at the future $pp$ collider ($50-100$TeV)
\cite{FCC}. Thus, we consider a sample effective theory with $\,\cut = \order{50\,\text{TeV}}\,$.\,
The relevant $WW$ scattering energies at such a $pp$ collider will be in the range of $\,0.5-30\,$TeV.
In parallel with Fig.\,\ref{fig:5}, we present the $WW$ scattering cross sections
Fig.\,\ref{fig:6} with the scattering energy $\,E=0.5-30\,$TeV, where we input the Higgs-curvature
coupling, $\,\xi =(3\times\! 10^{14},\,1.5\times\! 10^{14},\,0)$,\, as depicted by
the (red,\,blue,\,black) curves, respectively.
We have added a simple cut $\,|\cos\theta|< 0.995\,$ in each plot to
remove the special kinematical region of $t/u$-channel diagrams around $\,\theta = 0,\,\pi\,$.\,
In Fig.\,\ref{fig:6}(a), we also plot the RG improved cross sections in dashed curves
for nonzero $\,\xi\,$,\, where we input the initial values
$\,\xi =(3\times\! 10^{14},\,1.5\times\! 10^{14})$ at
$\,E=30\,$TeV and then include the one-loop RG running effects as functions of the $WW$ scattering energy.
This shows that the RG effects of $\,\xi\,$ are still negligible over the energy range
$\,E\lesssim 30\,$TeV for $WW$ scatterings.
From Fig.\,\ref{fig:6}, we see that the sensitivity to probing the
$\xi$ coupling may be improved by about a factor of $10$ as compared to Fig.\,\ref{fig:5} for the LHC case.
This illustrates the importance of increasing the $pp$ collision energy
up to $\,50-100$\,TeV.
It is encouraging to further perform detailed Monte Carlo simulations of the full $WW$ scattering processes
at the LHC\,(14\,TeV) and the future $pp$ collider ($50-100$\,TeV), where the
signals/backgrounds from the full processes $pp\to jj VV$ ($V=W,Z$) with $VV$ decays
will be analyzed.  This is fully beyond the current scope and will be considered elsewhere.

\vspace*{2mm}
\section{Unitarity Analysis for Higgs Inflation}
\label{BkgD}
\vspace*{2mm}

In this section, we will extend Sec.\,\ref{JFanalysis}-\ref{sec3.3}
to the situation with a generically large background field $\,\bar\phi\,$,\,
and perform the analysis of perturbative unitarity for the Higgs inflation \cite{bezrukov}.
In Sec.\,\ref{sec:4.1}, we present the background-dependent formulation.
Then, in Sec.\,\ref{sec:4.2} we derive the weak boson scattering amplitudes for a
general background field $\,\bar\phi\,$.\,
Finally, in Sec.\,\ref{sec:4.3}
we quantitatively analyze the background-dependent unitarity constraints
on the parameter space of Higgs inflation for both the conventional Higgs inflation \cite{bezrukov}
and the improved models\,\cite{B2new1} in light of the recent BICEP2 data \cite{BICHEP2}.

\vspace*{2mm}
\subsection{Background-Dependent Formulation}
\label{sec:4.1}
\vspace*{2mm}

We start the analysis in Jordan frame, in parallel with Sec.\,\ref{JFanalysis}.
Since the background field $\,\phib\,$ will vary,
we need to define the background-dependent Planck mass,
$\,\overbar{M}_{\textrm{Pl}}^2\equiv M^2+\xi\bar\phi^2$\,.\,
Expanding the metric tensor
$\,g_{\mu\nu}^{(J)}=\bar{g}^{(J)}_{\mu\nu}+\bar\kappa\hat{h}_{\mu\nu}^{(J)}$\,
with $\,\bar\kappa\equiv\sqrt{2}/\overbar{M}_{\textrm{Pl}}$,\,
we have the following transformations for diagonalizing the kinetic terms,
\beqa
\label{eq:KinDiagBD}
\hat{h}_{\mu\nu}^{(J)} \,=\,
h_{\mu\nu}^{(J)}-\eta_{\mu\nu}^{}\xi\bar\kappa\bar\phi\bar\zeta\phi^{(J)},
\qquad
{\hat{\phi}} =\, \bar\zeta\phi^{(J)},
\eeqa
where $\,h_{\mu\nu}^{(J)}\,$ and $\,\phi^{(J)}\,$ represent canonical fields in Jordan frame.
The modified rescaling factor $\,\bar{\zeta}\,$ is given by
\begin{eqnarray}
\label{eq:zetabphi}
\bar\zeta \,\equiv\, (1+6\xi^2\bar\phi^2/\,\overbar{M}_{\textrm{Pl}}^2)^{-1/2} \,.
\end{eqnarray}
In Jordan frame, there is no rescaling of Goldstone field $\,\pi\,$,\,
and we have $\,\pi^{(J)}=\pi$\,.\,
Compared with Eqs.\,(\ref{eq:kineticnorm})-(\ref{rescalezeta}),
this amounts to the replacements,
$\,(v,\,\kappa,\,\zeta)\to (\phib,\,\bar\kappa,\,\bar\zeta\,)$\,.\,

Then, we make transformations from Jordan frame to Einstein frame.
The Weyl factor (\ref{Omega}) can be rewritten as
\begin{eqnarray}\label{bOmega}
\Omega^2 \,=\,
\overbar\Omega^2\left[1+\bar{q}
\left(\frac{2\hat\phi}{\bar\phi}+\frac{\hat\phi^2+{{|\pi|^2}}}{\bar\phi^2}\right)\right],
\end{eqnarray}
with $\,\overbar\Omega\equiv \overbar{M}_{\textrm{Pl}}^{}/M_{\textrm{Pl}}^{}\,$
and $\,\bar{q}\,$ given by
\begin{eqnarray}\label{bq}
\bar{q} \,\equiv\, \frac{\,\xi\bar\phi^2\,}{\,M_{\textrm{Pl}}^2\overbar\Omega^2\,}
\,=\, \frac{\,\xi\bar\phi^2\,}{\,\overbar{M}_{\textrm{Pl}}^2\,} \,.
\end{eqnarray}
The scalar kinetic terms become,
\begin{eqnarray}
\mathcal{L}_{\textrm{kin}}^{} \,=\,
\frac{1}{2\overbar\Omega^2}
\left(\!1+\frac{6\xi^2\bar\phi^2}{M_{\textrm{Pl}}^2\overbar\Omega^2}\right)\!
(\partial_\mu\hat\phi\,)^2 +\frac{1}{2\overbar\Omega^2}|\partial_\mu{{\pi}}|^2 .
\end{eqnarray}
Thus, we further define the canonical fields $\,(\phi^{(E)},\, \pi^{(E)} )\,$ in Einstein frame via,
$\,\hat\phi = \overbar\Omega\bar\zeta\phi^{(E)}$\, and
$\,{\pi} =\overbar\Omega\pi^{(E)}$,\,  with the same $\,\bar\zeta\,$ in (\ref{eq:zetabphi}).
Comparing the canonical fields in the two frames, we have,
$\,\overbar\Omega =\phi^{(J)}\!/\phi^{(E)} = \pi^{(J)}\!/\pi^{(E)}$,\,
which amounts to a scale transformation for the fields.

\vspace*{2mm}
\subsection{Analysis of Scattering Amplitudes in Large Field Background}
\label{sec:4.2}
\vspace*{2mm}

To study the amplitudes in large field background, we will generalize previous analysis
to include all the relevant higher order $\xi$-dependent terms at
$\,\order{E^2}\,$ under the $\,1/\Mp^{}\,$ expansion.
We find that the calculations in Einstein frame are much simpler than that in Jordan frame.
For the application to Higgs inflations in Sec.\,\ref{sec:4.3},
we will focus on the analysis in Einstein frame.
We will also comment on the case of Jordan frame in the end.

In Einstein frame, the analysis for large field background should include higher order terms
of $\,\xi\bar\phi^2/\Mp^2\,$ since they become $\,\mathcal{O}(1)\,$
at $\,\bar\phi\sim\Mp/\!\sqrt\xi$\,.\,
Thus, we generalize the formulation of Sec.\ref{formal}--\ref{WBscat}, and derive
the scattering amplitudes in a generic background $\,\bar\phi\,$.\,
By including the higher order contributions, the scalar interactions in (\ref{DL_int_ss}) become,
\begin{align}
\label{DL-int_ssG}
\Delta\ld_{\text{int}}^{ss}  \,=\,
& -\FR{q}{2v^2}\Big(|\pd_\mu\pi|^2+\zeta^2(\pd_\mu \phi)^2\Big)\!
   \Big(|\pi|^2+(1-4q)\zeta^2\phi^2+2v\zeta \phi\Big)
\n\\[1.5mm]
& +\FR{1-\zeta^2}{8\zeta^2}\Big(1-\frac{2q}{v^2}(|\pi|^2+\zeta^2 \phi^2
  +2v\zeta \phi)+\frac{12q^2}{v^2}\zeta^2\phi^2\Big)\!
  \Big[\pd_\mu\Big(|\pi|^2+\zeta^2 \phi^2+2v\zeta \phi\Big)\Big]^2,~~~
\end{align}
where $\,q\equiv \xi v^2/\Mp^2$\,,\,
and we have rewritten the vertex coefficient $\,\xi^2v^2/\Mp^2\,$ in terms of $\,\zeta^2$\,.\,
In Eq.\,\eqref{DL-int_ssG}, we only consider the operators involving cubic or quartic vertices.
The Higgs-gauge interactions in (\ref{DL-int_pg}) are replaced by
\begin{align}
  \label{DL-int_pgG}
  \Delta\ld_{\text{int}}^{\phi g}
  \,=\, \big(2m_W^2 W_\mu^+W^{\mu -} \!+ m_Z^2 Z_\mu^2\big)
  \bigg[\Big(1\!-q\Big)\FR{\zeta}{v}\phi + \Big(1\!-5q+4q^2\Big)\FR{\zeta^2}{2v^2}\phi^2\bigg].
  ~~~~~
\end{align}
With these, we deduce general Feynman rules for the gauge and Goldstone boson scatterings.
Derivation for pure Goldstone scattering amplitudes uses the same Feynman diagrams,
while the Goldstone exchange should be further included for the processes with external Higgs bosons.
As a generalization of Eq.\,(\ref{GoldAmp}), we rederive the full amplitudes at $\,\mathcal{O}(E^2)$\,
and for all electrically neutral channels,
\beqa
\label{qGoldAmpGeneral}
  \T(\pi^+\pi^- \!\to \pi^+\pi^-) &\,=\,&
  \[1-(1-q)^2\zeta^2\](1\!+\cos\theta)\FR{\,E^2\,}{\,2v^2\,}\,,
\n\\[1mm]
  \T(\pi^+\pi^- \!\to \pi^0\pi^0) &\,=\,&
  \[1-(1-q)^2\zeta^2\]\FR{E^2}{v^2}\,,
\n\\[1mm]
  \T(\pi^+\pi^- \!\to \phi\phi) &\,=\,&
  (1-q)\left[3q+(1-\zeta^2)(1-2q)-q\zeta^2\right]\!\zeta^2\FR{E^2}{v^2}\,,
\\[1mm]
  \T(\pi^0\pi^0 \!\to \phi\phi) &\,=\,&
  (1-q)\left[3q+(1-\zeta^2)(1-2q)-q\zeta^2\right]\!\zeta^2\FR{E^2}{v^2}\,,
\n\\[1mm]
\T(\pi^0 \phi \to \pi^0 \phi) &\,=\,&
  -(1-q)\left[3q+(1-\zeta^2)(1-2q)-q\zeta^2\right]\!(1\!-\cos\theta)\zeta^2\FR{E^2}{2v^2} \,.
\hspace*{15mm}
\n
\eeqa
There are two elastic channels having no contributions at the $\,\order{E^2}\,$,\, i.e.,
$\,\T(\pi^0\pi^0\to \pi^0\pi^0) \simeq \T(\phi\phi\to\phi\phi) =  \order{E^0}\,$.\,
Using the above generalized amplitudes, we have also verified the validity of equivalence theorem,
which serves as nontrivial consistency checks of our calculation.

Next, we analyze the scattering amplitudes in a generic field background $\,\phib\,$.\,
Note that for Einstein frame $\,v\,$ denotes VEV of $\,\phi^{(E)}$\, in the electroweak vacuum.
Considering the rescaling factor $\,\overbar\Omega\,$ in (\ref{bOmega}),
we can derive the new amplitudes from  (\ref{qGoldAmpGeneral}) via the replacements
$\,v\to \bar\phi/\,\overbar\Omega$,\, $\zeta\to\bar\zeta$\,
and $\,q\to \bar{q}\,$ [cf.\ Eqs.\,(\ref{eq:zetabphi})-(\ref{bq})].
For the following sample processes, we have
\beqs
\label{EFTbkgd0}
\begin{eqnarray}
\mathcal{T}(\pi^+\pi^- \!\to\pi^0\pi^0)
&\,=\,&
\[1-(1-\bar{q})^2\bar\zeta^2\]\,\overbar\Omega^2\frac{E^2}{\bar\phi^2} \,,
\\
\mathcal{T}(\pi^+\pi^- \!\to\phi\phi)
&\,=\,&
(1-\bar{q})\[3\bar{q}+(1-2\bar{q})(1-\bar\zeta^2)-\bar{q}\bar\zeta^2\]
\bar\zeta^2\overbar\Omega^2\frac{E^2}{\bar\phi^2} \,.
\end{eqnarray}
\eeqs
In the limit that $\,\bar\phi\ll\Mp/\xi$\,,\, we have
\,$(\bar{q},\, \bar\zeta,\, \overbar\Omega)$\, reduce to the quantities
$(q,\, \zeta,\, 1)$ as defined in Sec.\,\ref{formal}--\ref{WBscat}.
But, the inflation epoch has $\,\bar\phi\gtrsim\Mp/\sqrt\xi\,$,\,
and thus corresponds to the different limits
$\,\bar{q}\sim 1\,$ and $\,\bar\zeta\ll 1\,$ (for $\xi\gg 1$).\,
This means that the previous expansion under $\,q\ll 1$\, and $\,\zeta\sim 1\,$ no longer applies.
In Einstein frame, for large $\,\phib\,$ and $\,\xi\gg 1$,\, the exchange of graviton
is always suppressed by $\,\bar{\zeta}^2\,$ as compared to the contributions of (\ref{EFTbkgd0}).
For the sake of unitarity analysis in the background field $\,\bar\phi<\Mp\,$,\,
the graviton exchange is negligible.

Finally, we comment on the calculation in Jordan frame.
For the scattering processes containing no Higgs boson in the in/out states,
the calculation is straightforward. For instance, up to the constant rescaling for energy scale,
the amplitude of $\,W^+_LW^-_L\to Z_L^0Z_L^0$\, is the same as that in Einstein frame.
At $\,\mathcal{O}(E^2)$,\, with negligible contribution from graviton exchange, the diagrams and Feynman vertices are the same. For the corresponding Goldstone scattering $\,\pi^+\pi^-\to\pi^0\pi^0$,\,
we find that $\,\xi\neq 0\,$ contribution from graviton exchange equals
that of contact interaction in Einstein frame.
But there is a complication for scattering processes with external Higgs bosons.
For both the cubic and quartic couplings of the Higgs and Goldstone bosons,
the second order perturbative expansion of $\,\sqrt{-g\,}\mathcal{R}\,$
is needed to derive higher order terms in $\,\bar{q}\,$ and $\,\bar\zeta^2$\,.\,
This is more tedious than the first order expansion.
In contrast, the Einstein frame has this complication transformed into higher dimensional operators
of Higgs fields which are much easier to handle. Hence, we will perform the unitarity analysis for
Higgs inflations in the Einstein frame.

\vspace*{2mm}
\subsection{Unitarity Constraints for Higgs Inflation}
\label{sec:4.3}
\vspace*{2mm}

Now we use results of previous subsection to derive the perturbative unitarity bound for the Higgs inflation.
In the conventional Higgs inflation model, the Higgs field couples to Ricci scalar through the nonminimal
coupling (\ref{eq:NMC}) and plays the role of inflaton.
During the inflation epoch, the background value of the unnormalized Higgs field $\,\hat\phi\,$
can be around $\,\Mp/\!\sqrt{\xi}$\,,\, and the canonically normalized Higgs field $\,\phi\,$ will
reach $\,\order{\Mp}\,$.\, In this case, the Higgs potential $\,U\,$ in Einstein frame
becomes exponentially flat when expressed in terms of canonically normalized background field $\,\chi\,$,
\beqa
  U ~\simeq~ \FR{\lam \Mp^4}{4\xi^2}\Big(1-e^{-\frac{2\chi}{\sqrt{6} \Mp}}\Big)^2 \,.
\eeqa
At the classical level, the Planck normalization
$\,(U/\ep)^{1/4} = 0.0276\Mp$\, \cite{planck} requires $\,\xi\sim10^4$\, for
$\,\lam = \order{0.1}\,$,\, where $\,\ep\,$ is the first slow-roll parameter
and is related to the tensor-to-scalar ratio $\,r=16\,\ep\,$.\,
The conventional Higgs inflation\,\cite{bezrukov} has $\,\ep = r/16 \sim 1/N^2$\, with $\,N\,$
denoting the number of $e$-foldings.
Thus, the typical choice of $e$-folding number $\,N=50-60\,$
implies a fairly small $\,\ep$\, and $\,r\,$,\, which agrees with the Planck data \cite{planck}.
The inflation scale is defined as
$\,\Lambda_{\textrm{INF}}^{}=U^{1/4}\simeq \lambda^{1/4}\Mp/\!\sqrt\xi$\,.\,
Since the Higgs self-coupling $\,\lambda = \mathcal{O}(0.1)$\,,\,
it is characterized by $\,\Lambda_{\textrm{INF}}^{}\sim \Mp/\sqrt\xi$\,.\,

On the other hand, the recent BICEP2 observation\,\cite{BICHEP2}
on the large scale B-mode of CMB polarization suggests a rather large tensor-to-scalar ratio,
$\,r=0.20^{+0.07}_{-0.05}$\,.\, The BICEP2 results (if confirmed) will have important impact on
Higgs inflation.  While the typical parameter space of Higgs inflation predicts a quite small
$\,r\,$ as mentioned above, it is still possible to accommodate a sizable $\,r$\,,\,
as discussed in \cite{B2new1,B2new2,B2new3},
through the tuning of top quark mass or proper extensions of the model.
Due to the renormalization group running, the Higgs self-coupling $\,\lam\,$
becomes extremely small at the inflation scale, at which a not-so-flat region
of the Higgs potential is realized. Then, only a mildly large $\,\xi \sim 10\,$ is needed to
further flatten the potential. Since the shape of the potential is no longer exponential,
the tensor-to-scalar ratio $\,r\,$ can be sizable.
This scenario uses the Einstein-frame Higgs field to set the renormalization scale.
Alternatively, one may use the Jordan-frame Higgs field, and this approach can result
in a quadratic chaotic inflation in Einstein frame \cite{B2new1}.
In this case, the nonminimal coupling is required to be $\,\xi\sim 100\,$.\,
We note that for either case, the inflation scale
is generally fixed by inputting the BICEP2 data on $\,r\,$
via $\,\epsilon=r/16\,$.\, Thus, we have,
$\,\Lambda_{\textrm{INF}} = U^{1/4}\simeq 2.3\times 10^{16}$\,GeV,
which coincides with the conventional grand unification (GUT) scale.
Hence, $\,\Lambda_{\textrm{INF}}\,$ is much lower than $\,\Mp/\!\sqrt\xi\,$
for $\,\xi =\order{10-100}\,$.\,

In all cases above, we see that the nonminimal coupling in Higgs inflations is significantly larger
than one, $\,\xi>1$.
Thus, a typical inflation scale such as $\,\Mp/\!\sqrt{\xi}$\, will be apparently higher than
our unitarity bound of $\,\order{\Mp/\xi}\,$ as given in Eq.\,\eqref{eq:UB-xi} [Sec.\,\ref{sec3.3}],
and we may worry about the unitarity issue of perturbative analysis of Higgs inflation.
But, as discussed above, the background value of Higgs field during inflation is very large.
This means that we should rederive the unitarity bound for scattering amplitudes with large background,
rather than using the result of Sec.\,\ref{sec3.3}.
Since we have systematically derived the scattering amplitudes with large background in the previous
subsection, we are now ready to analyze the background-dependent unitarity constraints
on the Higgs inflation models.
The use of flat-spacetime amplitudes in nearly de Sitter background during inflation is justified,
because the scale of spacetime curvature
(as characterized by the Hubble parameter $H\sim 10^{14}$\,GeV)
is smaller than the inflation scale $10^{16}$\,GeV (as inferred from the energy density)
by two orders of magnitude, according to the Friedmann equation.

Let us first study the unitarity bounds for the representative scattering channels.
We impose $\,|\mathfrak{Re}\,a_{0}^{}|<1/2$\,  for the amplitudes in (\ref{EFTbkgd0})
after proper normalization of their initial/final states. Thus, we compute the unitarity limits
$\,\Lambda^{(E)}_{\pi\pi}\,$ and $\,\Lambda^{(E)}_{\phi\phi}\,$ for
$\,\pi^+\pi^-\to \pi^0\pi^0\,$ and $\,\pi^+\pi^-\to \phi\phi\,$, respectively,
\beqs
\label{EFBDUB}
\begin{eqnarray}
\label{eq:UB-Epipi}
\Lambda^{(E)}_{\pi\pi} &\,=\,&
\(\!\frac{8\sqrt{2}\,\pi}{\,1-(1-\bar{q})^2\bar\zeta^2\,}\!\)^{\!\!\frac{1}{2}}
\!\frac{\bar\phi}{\,\overbar\Omega\,} \,,
\\
\label{eq:UB-Ephiphi}
\Lambda^{(E)}_{\phi\phi} &\,=\,&
\(\!\frac{8\sqrt{2}\,\pi}
  {\,(1-\bar{q})\[3\bar{q}+(1-2\bar{q})(1-\bar\zeta^2)-\bar{q}\bar\zeta^2\]\,}\!\)^{\!\!\frac{1}{2}}
\!\frac{\bar\phi}{\,\bar\zeta\,\overbar\Omega\,} .
\end{eqnarray}
\eeqs
Considering the LHC constraint $\,|\xi| < (2.7-5.2)\times 10^{15}\,$ from Fig.\,\ref{fig:4},
we have $\,\xi v^2\ll M_{\textrm{Pl}}^2$\,,\,
and thus $\,M^2\simeq M_{\textrm{Pl}}^2$\,.\,
This gives, $\,\overbar{M}_{\textrm{Pl}}^2\simeq M_{\textrm{Pl}}^2+\xi\bar\phi^2\,$.\,
Hence, for $\,\xi\gg 1\,$,\, we can derive the following asymptotic behaviors,
\begin{eqnarray}
\Lambda^{(E)}_{\pi\pi} \,\sim\, \left\{\!\!
\begin{array}{cl}
\frac{M_{\textrm{Pl}}}{\xi},~~ \quad & \bar\phi\ll\frac{M_{\textrm{Pl}}}{\xi} ,
\\[2mm]
\bar\phi\,,~~ \quad & \frac{M_{\textrm{Pl}}}{\xi}\ll\bar\phi\ll\frac{M_{\textrm{Pl}}}{\sqrt{\xi}\,},
\\[2mm]
\frac{M_{\textrm{Pl}}^{}}{\sqrt{\xi}\,},~~
\quad & \bar\phi\gg\frac{M_{\textrm{Pl}}}{\sqrt{\xi}\,} ,
\end{array}\qquad
\right.
\Lambda^{(E)}_{\phi\phi} \,\sim\, \left\{\!\!
\begin{array}{cl}
\frac{M_{\textrm{Pl}}}{\xi},~~ \quad & \bar\phi\ll\frac{M_{\textrm{Pl}}}{\xi} ,
\\[3mm]
\frac{\,\xi\phib^2\,}{\Mp},~~
\quad & \frac{M_{\textrm{Pl}}}{\xi}\ll\bar\phi\ll\frac{M_{\textrm{Pl}}}{\sqrt\xi} ,
\\[2mm]
\sqrt{\xi}\phib,~~ \quad & \bar\phi\gg\frac{M_{\textrm{Pl}}}{\sqrt\xi} .
\end{array}
\right.
\hspace*{8mm}
\end{eqnarray}
For completeness, we also remark that
in the case of\,\footnote{The limit $\,M^2\to 0\,$ leads to $\,\Mp^2 = \xi v^2$.\,
This corresponds to the early scenarios of induced gravity inflation \cite{ING}.}
$\,M^2\ll M_{\textrm{Pl}}^2\,$,\,
we would have $\,\Lambda^{(E)}_{\pi\pi}\sim M_{\textrm{Pl}}/\sqrt\xi$\, and
$\,\Lambda^{(E)}_{\phi\phi} \sim (\Mp /M)\sqrt{\xi}\phib \gg \sqrt{\xi}\phib$\,.

\begin{figure}
\begin{center}
\includegraphics[width=0.55\textwidth]{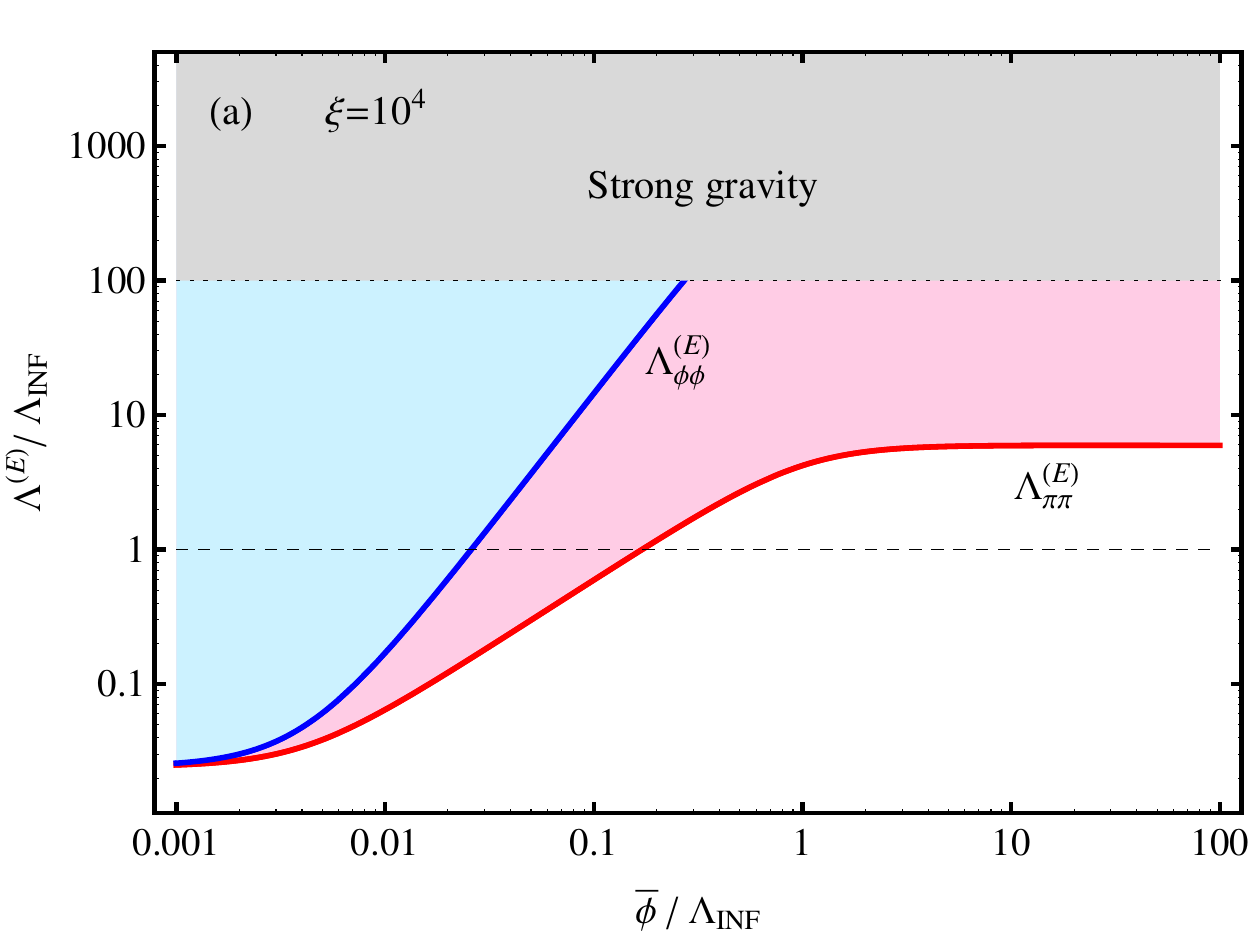}
\\
\includegraphics[width=0.49\textwidth]{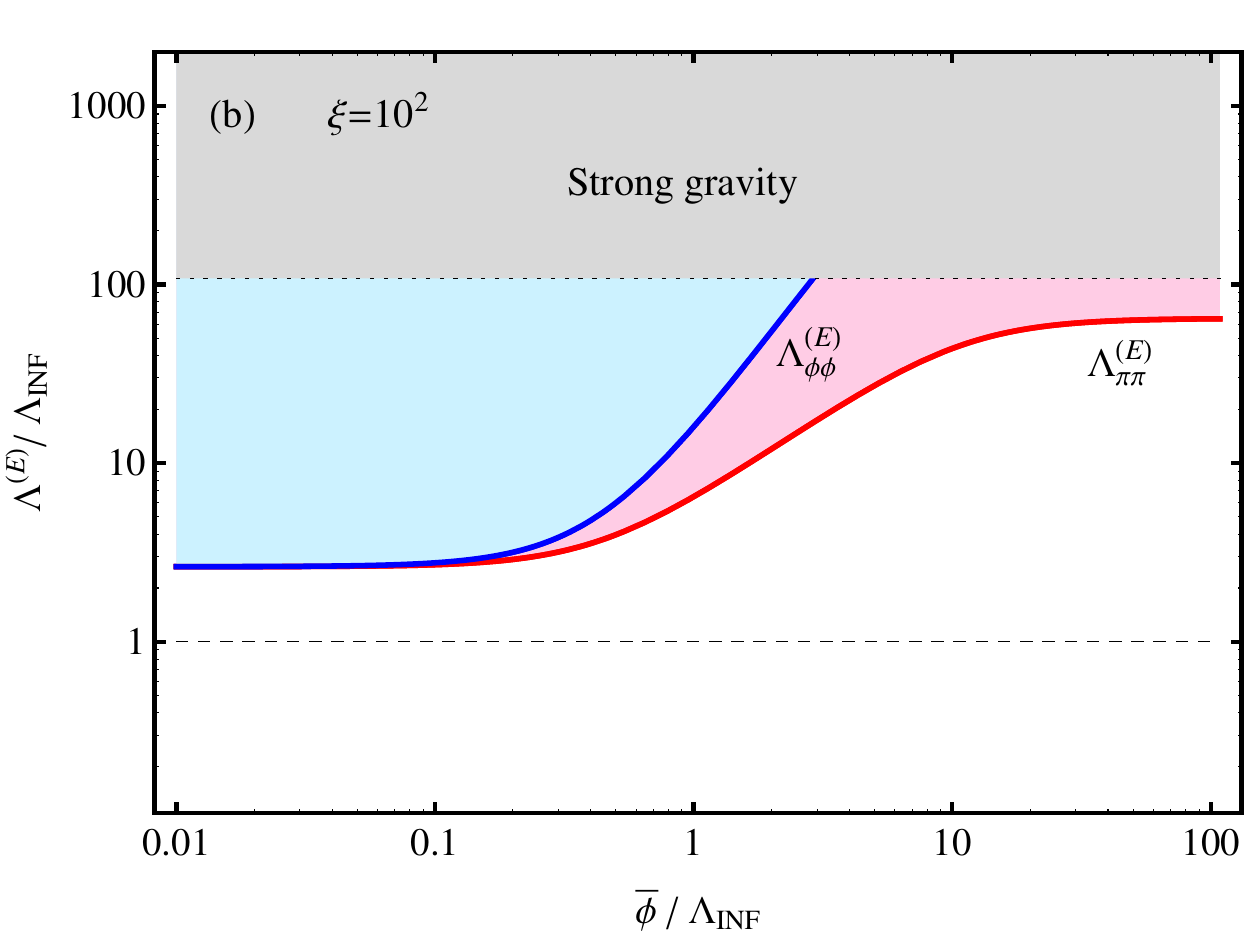}~~
\includegraphics[width=0.49\textwidth]{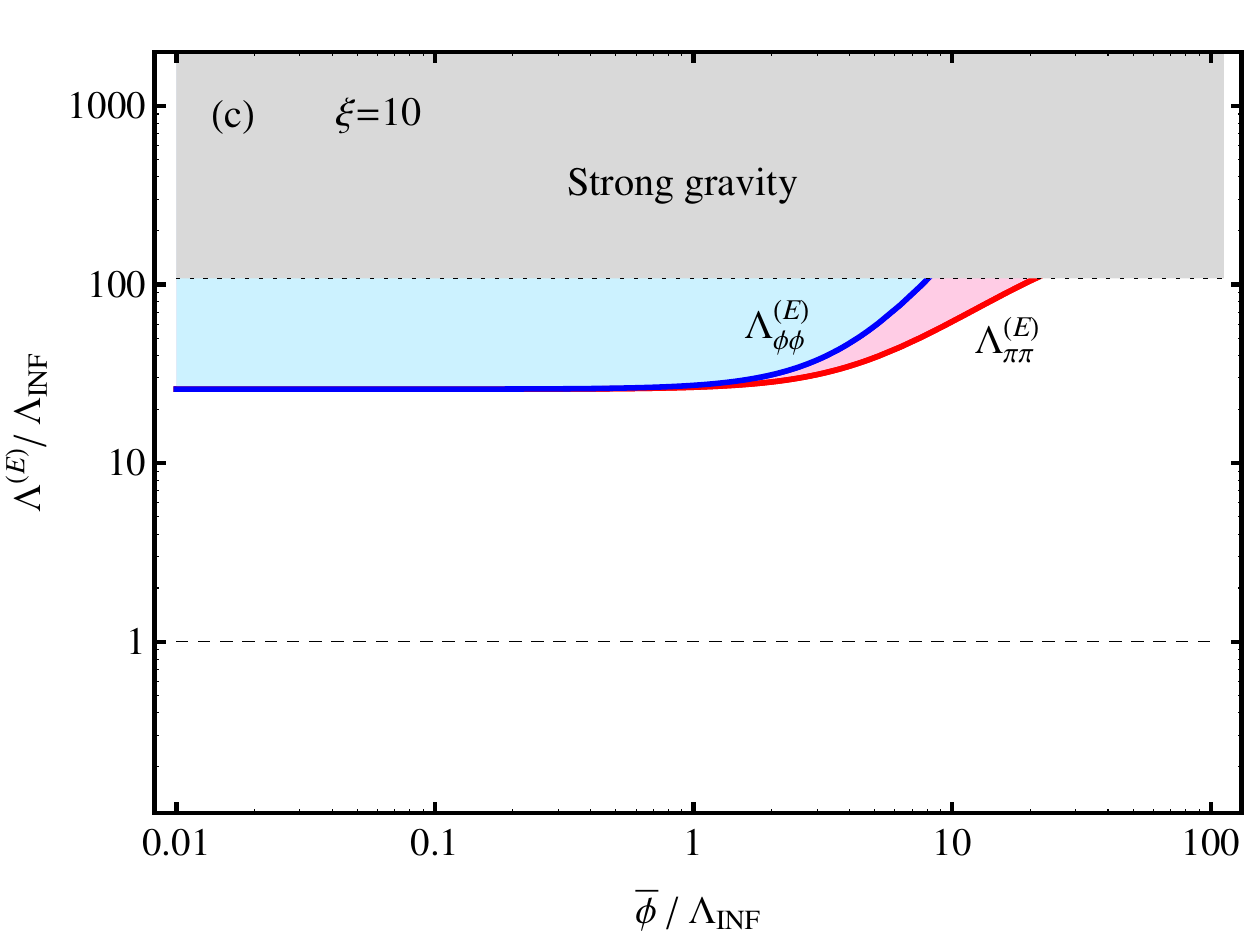}
\vspace*{-5mm}
\caption{Perturbative unitarity bounds $\Lambda^{(E)}$ as functions of $\,\phib\,$,\,
with $\,\xi=10^4\,$ [plot-(a)], $\,\xi=10^2\,$ [plot-(b)] and $\,\xi=10\,$ [plot-(c)].
Here, both axes are normalized by $\,\Lambda_{\textrm{INF}}^{}$\,.\,
In all plots, the blue and red curves denote $\,\Lambda_{\phi\phi}^{}\,$ and
$\,\Lambda_{\pi\pi}^{}$,\, respectively,
where the region above each curve violates perturbative unitarity.
The grey area with dotted boundary denotes the strong gravity region, which is
above the reduced Planck mass $\,\Mp\,$. The horizontal dashed line denotes the inflation scale $\,\Lambda_{\textrm{INF}}^{}\simeq\Mp/\!\sqrt\xi\,$ in plot-(a),
and $\,\Lambda_{\textrm{INF}}^{}\simeq 2.3\times 10^{16}\,$GeV\, in plot-(b,c)
as indicated by the BICEP2 data.}
\label{fig:7}
\end{center}
\end{figure}

To study the the unitarity constraints on Higgs inflation,
we present in Fig.\,\ref{fig:7} the unitarity bound
$\Lambda^{(E)}$\, as a function of $\,\bar\phi\,$,\,
with sample inputs $\,\xi =10^4\,$ [plot-(a)], $\,\xi =10^2\,$ [plot-(b)],
and $\,\xi =10\,$ [plot-(c)].
For clarity, we normalize both axes by $\,\Lambda_{\textrm{INF}}^{}$\,
in each plot. The red and blue curves denote the unitarity bounds $\,\Lambda^{(E)}_{\pi\pi}\,$
and $\,\Lambda^{(E)}_{\phi\phi}\,$,\, respectively.
The light pink and blue areas above each curve violate perturbative unitarity.
Plot-(a) demonstrates the conventional Higgs inflation\,\cite{bezrukov} with a large
$\,\xi =10^4\,$.
The inflation scale $\,\Lambda_{\textrm{INF}}\simeq\Mp/\!\sqrt\xi$\,
is depicted as the horizontal dashed line.
It shows that the strongest unitarity bound is higher than the effective inflation scale
$\,\Mp^{}/\!\sqrt{\xi}\,$\, for $\,\phib \gtrsim 0.2\Lambda_{\textrm{INF}}^{}\,$.\,
This agrees well with the schematic picture from recent qualitative estimates \cite{BgdDep}.
Plots (b) and (c) represent the improved models\,\cite{B2new1,B2new2,B2new3}
compatible with a large $\,r\,$ as indicated by the BICEP2 observation \cite{BICHEP2}.
For illustration, we set $\,\xi=10^2\,$ and $\,\xi=10\,$
for the two sample scenarios discussed above.
Here, the inflation scale $\,\Lambda_{\textrm{INF}}\simeq 2.3\times 10^{16}\,$GeV
is denoted by the horizontal dashed line at the ratio
$\,\Lambda^{(E)}\!/\Lambda_{\textrm{INF}}=1\,$.\,
It is lower than the unitarity bound even for the small background field regions.
This shows that the unitarity constraints are largely relieved for moderate values of
the nonminimal coupling, $\,\xi =\order{10-100}\,$.\,

In parallel with the perturbative unitarity bound \eqref{eq:UB-xi} derived
in the electroweak vacuum in Sec.\,\ref{sec3.3}, we perform a coupled channel analysis
for the large field background $\,\phib \gtrsim \Mp^{}/\!\sqrt{\xi}\,$.\,
For the four neutral channels,
$\,|\pi^+\pi^-\ra$, $\fr{1}{\sqrt 2}|\pi^0\pi^0\ra$, $\fr{1}{\sqrt 2}|\phi\phi\ra$,
and $|\pi^0 \phi\ra$,\,  we deduce the $s$-wave amplitudes,
\begin{align}
  a_0^{}(E) ~=&~
  \FR{\,\[1\!-(1\!-\bar{q})^2\bar\zeta^2\]\overbar\Omega^2E^2\,}{32\pi\,\phib^2}\!
  \begin{pmatrix}
    1 & \sqrt{2} & \sqrt{2}A & 0 \\[1mm]
    \sqrt{2} & 0 & A & 0 \\[1mm]
    \sqrt{2}A & A & 0 & 0 \\[1mm]
    0 & 0 & 0 & -A
  \end{pmatrix}
  \n\\[3mm]
  \simeq &~ \FR{\,\overbar\Omega^2E^2\,}{\,32\pi\bar\phi^2\,}\!
  \begin{pmatrix}
    1 & \sqrt2 & 0 & 0 \\[1mm]
    \sqrt2 & 0 & 0 & 0 \\
     0 & 0 & 0 & 0 \\
     0 & 0 & 0 & 0
  \end{pmatrix} \!.
\label{a0-4x4LB}
\end{align}
In the above, we have denoted,
\beqa
A ~=~ \frac{\,1\!+\bar{q}-(1\!-\bar{q})\bar\zeta^2\,}
           {1\!-(1\!-\bar{q})^2\bar\zeta^2}\frac{\bar\zeta^2}{\,\overbar\Omega^2\,}\,.
\eeqa
For the large background field $\,\phib\sim\Mp^{}/\!\sqrt{\xi}\,$ and large $\,\xi\gg 1\,$,\,
we have the asymptotic behavior $\,A\sim \xi^{-1}\ll 1\,$,\, and thus the $\,A\,$ term is
negligible for the current analysis.
This means that the processes with Higgs bosons in the external states have negligible contribution.
Imposing the unitarity condition on the largest eigenvalue of $\,a_0^{}\,$
via the above coupled channel analysis,
we infer the unitarity bound for large background field $\,\phib$\,,\,
\beqa
E ~<~ 2^{-\frac{1}{4}}_{}\Lambda_{\pi\pi}^{(E)} \,, 
\eeqa
where $\,\Lambda_{\pi\pi}^{(E)}\,$
is given by \eqref{eq:UB-Epipi}.  For comparison, we present this bound
in the $\,(E,\, |\xi|)\,$ plane at $\,\phib = \Mp^{}/\!\sqrt{\xi}$\,
in Fig.\,\ref{fig:4}(b) by the red curve, where the yellow region
violates perturbative unitarity.
It shows that the unitarity bound is significantly relieved when derived in the large field background,
which applies to the case of Higgs inflation.
In the same plot-(b) of Fig.\,\ref{fig:4}, we draw a vertical dashed line to depict the
inflation scale, $\,\Lambda_{\text{INF}}^{}\simeq 2.3\times 10^{16}\,$GeV,
as indicated by the BICEP2 data\,\cite{BICHEP2}.
For the scattering energy around the inflation scale,
$\,E \sim \Lambda_{\text{INF}}^{}$\,,\,
we see that the red curve imposes a unitarity bound,
$\,\xi < \order{10^{5-6}}\,$.\,

Finally, for comparison with the literature, we note that the interesting papers\,\cite{BgdDep}
discussed two types of unitarity bounds for both frames by simple power counting analysis.
It estimated the unitarity bound $\,\Lambda_{\textrm{g-s}}^{}$\, for scalar-gravity coupling of
$\,\phi-\phi-h_{\mu\nu}^{}$\, in the Jordan frame,
which contributes to the process $\,\phi\phi\to\phi\phi$\,.\,
As we showed earlier, this specific scattering amplitude has vanishing $\,E^2\,$ term
due to the crossing symmetry of $(s,\,t,\,u)$ channels.
It also estimated the unitarity bound $\,\Lambda_{\textrm{gauge}}^{}\,$ from
the gauge boson scattering.
These estimates\,\cite{BgdDep} agree with the main feature of our quantitative unitarity bounds
$\,\Lambda_{\pi\pi}^{}$\, in (\ref{EFBDUB}) and Fig.\,\ref{fig:7}.
As shown above, the $\,\Lambda_{\pi\pi}^{(E)}$\, places the best unitarity constraint over
the large background field region of the Higgs inflation.
Recently, Ref.\,\cite{Burgess} studied the impact of UV physics on the prediction of Higgs inflation.
They utilized the non-linear realization to discuss the unitarity of Higgs inflation, and
derived the leading order Goldstone amplitudes, which are in qualitative agreement with our results.
We have also derived the scattering amplitude with Higgs bosons as external states.
Furthermore, we presented the quantitative unitarity bounds
on the Higgs inflation models in Fig.\ref{fig:7}(a)-(c).

\section{Conclusions}
\label{conclusion}
\vspace*{2mm}

It is striking that the gravitational force not only shapes the world
at its macroscopic and cosmological scales,
but will also play key role at the fundamental Planck scale.
We would then ask: {\it what happens in between?}  Given the LHC discovery of a 125\,GeV Higgs boson
\cite{LHC2012,LHC2013}, it is strongly motivated for us to explore the Higgs gravitational
interactions in connection with the electroweak symmetry breaking mechanism
and the origin of inertial mass generation for all elementary particles,
as well as the Higgs inflation.

Combining the SM with general relativity (GR) as a joint effective theory, we note that there is
a unique dimension-4 operator (\ref{eq:NMC}) for the Higgs-gravity interactions
with nonminimal coupling $\,\xi\,$.
This provides a generic Higgs portal to the new physics beyond SM.
In this work, we systematically studied the contributions of this Higgs-gravity interaction
(\ref{eq:NMC}) to weak boson scattering processes in both Jordan and Einstein frames,
over the energy regions accessible by the LHC\,(14\,TeV) and the future circular $pp$ colliders
($50-100$\,TeV). We explicitly demonstrated the equivalence theorem in the presence of
Higgs-gravity coupling $\,\xi\,$ in both Jordan and Einstein frames.
For the $\xi$-induced leading amplitudes, we derived the full results at $\order{E^2}$,\,
which are needed for studying the case of large background field in Higgs inflations.
Then, we analyzed the perturbative unitarity bound on $\,\xi\,$ via coupled channel analysis
in the background of the electroweak vacuum.
We also verified the equivalence between the two frames
for computing the scattering amplitudes and cross sections.
This systematically extends our previous short study\,\cite{XRH}
with analysis in the Einstein frame alone
and only to the first order of $\,1/\Mp^{2}\,$.\,
For applications to Higgs inflation,
we further studied the weak boson scatterings and unitarity constraints for the large background
field case. We quantitatively established the viable perturbative parameter space
of the conventional Higgs inflation\,\cite{bezrukov} and the improved models\,\cite{B2new1,B2new2}
in light of the recent BICEP2 data \cite{BICHEP2}.

To be concrete, in Sec.\,\ref{formal} we presented the formulation in Jordan and Einstein frames.
We derived the $\xi$-induced Higgs-gravity interactions for both frames, and summarized
all the relevant Feynman rules in  Appendix\,\ref{A:FR}.
Then, in Sec.\,\ref{WBscat} we systematically analyzed longitudinal weak boson scattering
and the corresponding Goldstone boson scattering in both frames. In each frame,
we explicitly demonstrated the longitudinal-Goldstone boson equivalence theorem
with nonzero Higgs-gravity coupling $\,\xi\,$.\,
We further verified the equivalence between the two frames for all scattering processes.
In Sec.\,\ref{sec3.3}, we performed a coupled channel analysis of
weak boson scattering in the electroweak vacuum,
and derived unitarity bound on $\,\xi\,$ in Fig.\,\ref{fig:4}.
We further studied two intriguing scenarios, in which the UV cutoff for the SM\,+\,GR
effective theory is around $\,\cut = \order{10\,\text{TeV}}\,$ and
$\,\cut = \order{50\,\text{TeV}}\,$,\, respectively.
Thus, the $\,\xi\,$ coupling can reach up to $\,\xi =\order{10^{15}}\,$
for $\,\cut = \order{10\,\text{TeV}}$,\, or,
reach up to $\,\xi =\order{10^{14}}\,$
for $\,\cut = \order{50\,\text{TeV}}$,\, as shown in Fig.\,\ref{fig:4}(a).
In Fig.\,\ref{fig:5}, we presented our predictions of the
$WW$ scattering cross sections with coupling $\,\xi=\order{10^{15}}\,$,\,
over the energy scale $\,E(WW)=0.2-4\,$TeV, which is accessible at the LHC\,(14\,TeV).
These exhibit {\it different behaviors} from the naive SM result ($\xi =0$),
and thus will be discriminated by the upcoming runs at the LHC\,(14\,TeV) with
higher integrated luminosity.
We further analyzed the $WW$ scattering cross sections in the energy range
of $\,E(WW)=1-30\,$TeV. These will be realized at the future circular $pp$
colliders ($50-100$\,TeV) \cite{FCC}, which may have sensitivity to probe the
Higgs-gravity coupling at the level of $\,\xi=\order{10^{14}}\,$,\,
as shown in Fig.\,\ref{fig:6}.
We suggest that the $\,\xi\,$ coupling can be further probed by invoking
the cubic Higgs self-interactions [Eq.\,\eqref{DL_int_ss}]
at the future high energy $pp$ colliders.

In Sec.\,\ref{BkgD}, we studied the Higgs-field-background dependent
weak boson scattering amplitudes, and quantitatively performed the unitarity analysis
for the Higgs inflation models.
We generalized the formulation of Sec.\,\ref{formal} to
a generic Higgs-field background in both Jordan and Einstein frames.
We derived the new Feynman rules and the scattering amplitudes accordingly.
For the case of large field background, we have taken account of the full contributions
at $\,\order{E^2}\,$ for the scattering amplitudes.  With these, we demonstrated
that the unitarity bound on the $\xi$ coupling is substantially relieved,
as shown in Fig.\,\ref{fig:4}(b). Finally, we applied this analysis to the conventional
Higgs inflation\,\cite{bezrukov} and the improved models\,\cite{B2new1,B2new2}
in light of the recent BICEP2 observation\,\cite{BICHEP2}.
We quantitatively analyzed the viable perturbative parameter space for the
Higgs inflation models, as shown in Fig.\;\ref{fig:7}(a)-(c)
for three sample inputs of the $\,\xi\,$ coupling.

\vspace*{5mm}
\noindent
{\bf Acknowledgements}\\[1.5mm]
We thank John R.\ Ellis for valuable discussions during his recent visit at Tsinghua HEP Center.
We thank Michael Trott and Xavier Calmet for valuable discussions on this subject.
This work was supported by National NSF of China (under grants 11275101, 11135003)
and National Basic Research Program (under grant 2010CB833000).


\newpage
\noindent
{\bf\Large Appendix}
\appendix

\section{Feynman Rules in Jordan and Einstein Frames}
\label{A:FR}
\vspace*{1mm}

In this appendix, we present all the relevant Feynman rules
at $\order{\Mp^{-2}}$ in the electroweak vacuum for evaluating
longitudinal and Goldstone boson scattering in both Jordan and Einstein frames.
We have retained the rescaling factor $\,\zeta\,$ for the Higgs field $\,\phi\,$
without expansion. This will also allow us to extract the leading terms of
$\,(\xi v/\Mp^{})^{2n}\,$ for the scattering amplitudes at $\,\order{E^2}\,$.\,
In our notation, all momenta will flow inward.

\vspace*{2mm}
\subsection{Feynman Rules in Jordan Frame}
\label{A:FRJ}
\vspace*{2mm}

{\allowdisplaybreaks

\beqa
\label{eq:A1-1}
  \parbox{32mm}{\includegraphics[width=32mm]{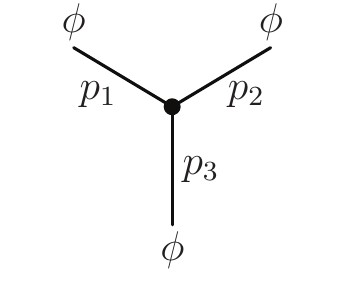}}
  &&
  \begin{split}
  =& -\ii 6\zeta^3\lam v
     +\zeta^3\!\(\!-\FR{\ii\xi v}{\Mp^2}+\FR{\ii 6\xi^2v}{\Mp^2}\!\)\!\big(p_1^2+p_2^2+p_3^2\big)\,.
  \end{split}
\\[2mm]
  \parbox{32mm}{\includegraphics[width=32mm]{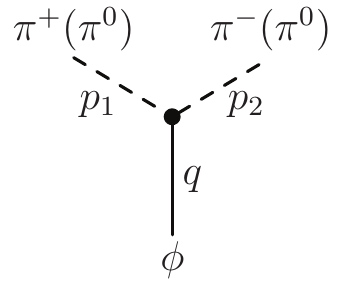}}
  &&= - \ii 2\lam\zeta v+\FR{\,\ii 2\xi\zeta v\,}{\Mp^2}(p_1\cdot p_2)+\FR{\,\ii 6\xi^2\zeta v\,}{\Mp^2} q^2,
\\[2mm]
  \parbox{32mm}{\includegraphics[width=32mm]{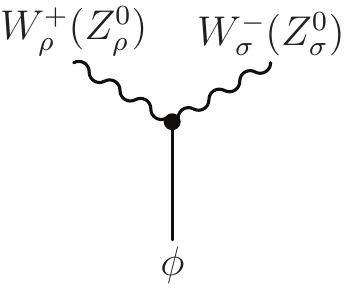}}
  &&=\FR{\ii 2 m_{W(Z)}^2}{v}\bigg(1-\FR{\xi v^2}{\Mp^2}\bigg)\zeta\eta^{\rh\si},
\\[2mm]
  \parbox{32mm}{\includegraphics[width=32mm]{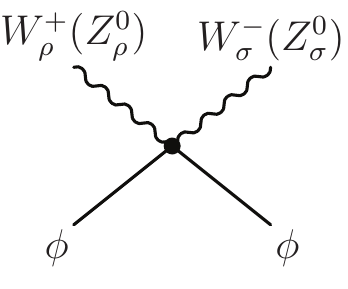}}
  &&=\FR{\ii 2 m_{W(Z)}^2}{v^2}\bigg(1-\FR{4\xi v^2}{\Mp^2}\bigg)\zeta^2\eta^{\rh\si},
\\[2mm]
  \parbox{32mm}{\includegraphics[width=32mm]{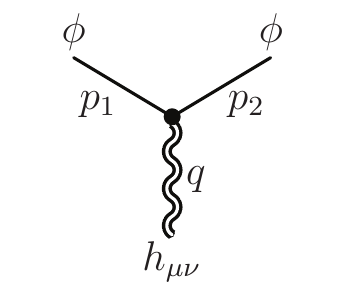}}
  &&\begin{split}
     =\FR{\,\ii\sqrt{2}\,}{\Mp}\zeta^2\Big[\xi(q^\mu q^\nu-q^2\eta^{\mu\nu})+\big(p_1^{(\mu}p_2^{\nu)}-\fr{1}{2}p_1\cdot p_2\eta^{\mu\nu}\big)\Big],
   \end{split}
\\[2mm]
  \parbox{32mm}{\includegraphics[width=32mm]{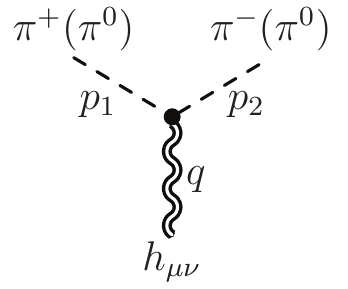}}
  &&\begin{split}
     =\FR{\,\ii\sqrt{2}\,}{\Mp} \Big[\xi(q^\mu q^\nu-q^2\eta^{\mu\nu})+\big(p_1^{(\mu}p_2^{\nu)}-\fr{1}{2}p_1\cdot p_2\eta^{\mu\nu}\big)\Big],
   \end{split}
\\[5mm]
  \parbox{32mm}{\includegraphics[width=32mm]{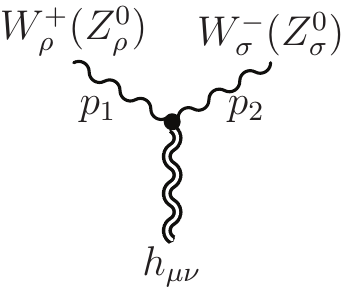}}
  &&\begin{split}
     = -\FR{\,\ii\sqrt{2}\,}{\Mp} \Big[&p_1^{(\mu}p_2^{\nu)}
       \eta^{\rh\si}+\fr{1}{2}p_1^{\si} p_2^{\rh}\eta^{\mu\nu}_{}-p_1^\si p_2^{(\nu}\eta^{\mu)\rh}_{}-p_2^\rh p_1^{(\mu}\eta^{\nu)\si}_{}
\\
    &+(p_1\cdot p_2+m_{W,Z}^2)\big(\eta^{\rh(\mu}\eta^{\nu)\si}-\fr{1}{2}\eta^{\mu\nu}\eta^{\rh\si}\big)\Big].
   \end{split}
\end{eqnarray}

}

\vspace*{5mm}
\subsection{Feynman Rules in Einstein Frame}
\label{EFfeynmanrule}
\vspace*{2mm}

{\allowdisplaybreaks

\begin{eqnarray}
  \parbox{32mm}{\includegraphics[width=32mm]{fig/FD_scalar_1.pdf}}
  &&\begin{split}
    =&  -\ii 6\zeta^3\lam v
    +\zeta^3\!\(\!-\FR{\ii\xi v}{\Mp^2}+\FR{\ii 6\xi^2v}{\Mp^2}\!\)\!\big(p_1^2+p_2^2+p_3^2\big)\,,
    \end{split}
\\[5mm]
  \parbox{32mm}{\includegraphics[width=32mm]{fig/FD_scalar_2.pdf}}
  &&= - \ii 2\lam\zeta v+\FR{\,\ii 2\xi\zeta v\,}{\Mp^2}(p_1\cdot p_2)+\FR{\,\ii 6\xi^2\zeta v\,}{\Mp^2} q^2,
\\[4mm]
  \parbox{32mm}{\includegraphics[width=32mm]{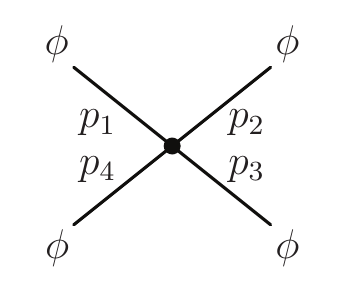}}
  &&\begin{split}
     = & -\ii 6\lam\zeta^4+\FR{\ii 2\xi\zeta^4}{\Mp^2}\big(p_1\cdot p_2+\text{5 permutations}\big)
       \\
       & +\FR{\ii 3\xi^2\zeta^4}{\Mp^2}\[(p_1\!+p_2)^2+\text{5 permutations}\] \!,
       \\
  \end{split}
\\[5mm]
  \parbox{32mm}{\includegraphics[width=32mm]{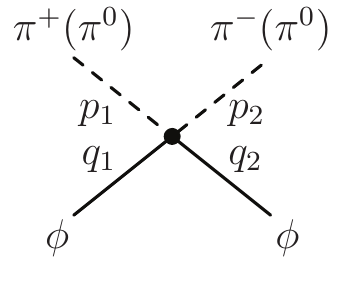}}
  &&=-\ii\lam\zeta^2+\FR{\ii 2\xi\zeta^2}{\Mp^2}(p_1\cdot p_2 + q_1\cdot q_2)+\FR{\ii 6\xi^2\zeta^2}{\Mp^2}(p_1+p_2)^2,
\\[4mm]
  \parbox{32mm}{\includegraphics[width=32mm]{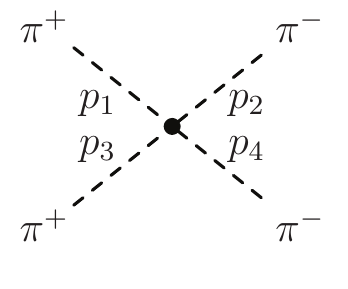}}
  &&\begin{split}
    = &
    -\ii 2\lam-\FR{\ii 2\xi}{\Mp^2}(p_1\!+p_3)^2
    +\FR{\ii 6\xi^2}{\Mp^2}\[ (p_1\!+p_2)^2+(p_1\!+p_4)^2\],
  \end{split}
\\[4mm]
  \parbox{32mm}{\includegraphics[width=32mm]{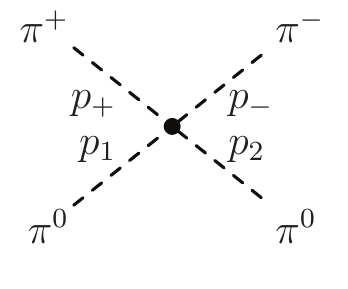}}
  && = -\ii 2\lam+\FR{2\ii\xi}{\Mp^2}(p_+\cdot p_-+p_1\cdot p_2)
       +\FR{\ii 6\xi^2}{\Mp^2}(p_++p_-)^2,
\\[5mm]
  \parbox{32mm}{\includegraphics[width=32mm]{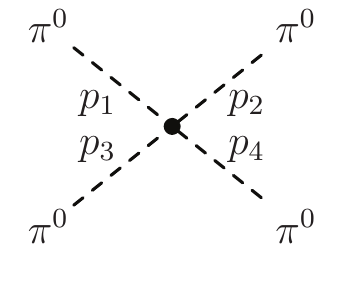}}
  &&\begin{split}
     = & -\ii 6\lam+\FR{\ii 2\xi}{\Mp^2}\big(p_1\cdot p_2+\text{5 permutations}\big)\\
       & +\FR{\ii 3\xi^2}{\Mp^2}\[(p_1\!+p_2)^2+\text{5 permutations}\] \!,\\
  \end{split}
\\[5mm]
  \parbox{32mm}{\includegraphics[width=32mm]{fig/FD_vector_8.pdf}}
  &&=\FR{\ii 2 m_{W(Z)}^2}{v}\bigg(1-\FR{\xi v^2}{\Mp^2}\bigg)\zeta\eta^{\rh\si},
\\[5mm]
  \parbox{32mm}{\includegraphics[width=32mm]{fig/FD_vector_9.pdf}}
  && =\FR{\ii 2m_{W(Z)}^2}{v^2}\bigg(1-\FR{5\xi v^2}{\Mp^2}\bigg)\zeta^2\eta^{\rh\si}.
\end{eqnarray}

}


\newpage

\end{document}